\documentclass[12pt]{article}
\usepackage{epsf,amsbsy}

\def\a{\alpha}
\def\b{\beta}
\def\C{\raise2pt\hbox{\rm\large$\chi$}}
\def\c{\chi}
\def\d{\delta}
\def\e{\epsilon}
\def\f{\phi}
\def\g{\gamma}
\def\h{\eta}

\def\k{\kappa}
\def\l{\lambda}
\def\m{\mu}
\def\n{\nu}
\def\o{\omega}
\def\p{\pi}
\def\q{\theta}
\def\r{\rho}
\def\s{\sigma}
\def\t{\tau}

\def\x{\xi}
\def\z{\zeta}
\def\D{\Delta}
\def\F{\Phi}

\def\J{\Psi}
\def\L{\Lambda}
\def\O{\Omega}
\def\P{\Pi}
\def\Q{\Theta}
\def\S{\Sigma}

\def\Bvf{\boldsymbol{\varphi}}
\def\Bc{\boldsymbol{\chi}}

\def\ve{\varepsilon}
\def\vf{\varphi}

\def\bz{\bar{z}}
\def\bt{\bar{\t}}
\def\bo{\bar{\o}}
\def\br{\bar{\r}}

\def\bC{\bar{C}}
\def\bT{\bar{T}}
\def\bO{\bar{\O}}
\def\bK{\bar{K}}
\def\baf{\bar{\f}}

\def\ha{\hat{\a}}

\def\hT{\hat{T}}
\def\hG{\hat{G}}
\def\hJ{\hat{J}}

\def\tb{\tilde{b}}
\def\tc{\tilde{c}}
\def\ta{\tilde{\a}}
\def\tha{\tilde{\ha}}
\def\tl{\tilde{l}}
\def\ts{\tilde{\s}}
\def\th{\tilde{\h}}
\def\te{\tilde{\e}}

\def\tA{\tilde{A}}
\def\tB{\tilde{B}}

\def\tm{\tilde{m}}
\def\tn{\tilde{n}}

\DeclareFontFamily{OT1}{msb}{}{}
\DeclareFontShape{OT1}{msb}{m}{n}
 {  <5> <6> <7> <8> <9> <10> gen * msbm
      <10.95><12><14.4><17.28><20.74><24.88>msbm10}{}
\DeclareMathAlphabet{\bubble}{OT1}{msb}{m}{n}

\def\bR{{\bubble R}}
\def\bZ{{\bubble Z}}
\def\buC{{\bubble C}}

\def\cd{{\cal D}}

\def\ck{{\cal K}}
\def\cl{{\cal L}}
\def\cm{{\cal M}}

\def\cp{{\cal P}}

\def\car{{\cal R}}

\def\cz{{\cal Z}}

\def\Im{{\rm Im}}
\def\Re{{\rm Re}}
\def\sdet{{\rm sdet}}
\def\res{{\rm res}}
\def\det{{\rm det}}
\def\PCO{P\!C\!O}
\def\SFO{S\!F\!O}

\def\pa{\partial}                                       

\def\dg{{}^{\dagger}}                                     

\def\ra{\rightarrow}

\def\ua{{\scriptscriptstyle \uparrow}}
\def\da{{\scriptscriptstyle \downarrow}}

\def\ket#1{\left| #1\right\rangle}              
\def\VEV#1{\left\langle #1\right\rangle}        
\def\abs#1{\left| #1\right|}                    
\def\leftrightarrowfill{$\mathsurround=0pt \mathord\leftarrow \mkern-6mu
        \cleaders\hbox{$\mkern-2mu \mathord- \mkern-2mu$}\hfill
        \mkern-6mu \mathord\rightarrow$}
\def\dvec#1{\vbox{\ialign{##\crcr
        \leftrightarrowfill\crcr\noalign{\kern-1pt\nointerlineskip}
        $\hfil\displaystyle{#1}\hfil$\crcr}}}           

\def\fr#1#2{{\textstyle{#1\over\vphantom2\smash{\raise.20ex
        \hbox{$\scriptstyle{#2}$}}}}}                   
\def\parvar#1#2{{\d #1\over \d #2}}               

\def\be#1{\b^{#1}}
\def\ga#1{\g^{#1}}
\def\Ge#1{G^{#1}}
\def\hGe#1{\hG^{#1}}
\def\fu#1{\vf^{#1}}

\def\ep#1#2{\e^{#1}_{#2}}
\def\tep#1#2{\te^{#1}_{#2}}
\def\bep#1{\bar{\e}^{#1}}
\def\et#1{\h^{#1}}
\def\xx#1{\x^{#1}}
\def\bet#1{\bar{\h}^{#1}}
\def\zu#1#2{Z^{{#1}#2}} 
\def\zd#1#2{Z^{{#1}}{}_#2} 
\def\pu#1#2#3{\J^{{#1}#3}_{{#2}}} 
\def\pd#1#2#3{\J^{{#1}}_{{#2}}{}_#3} 
\def\bpu#1#2#3{\bar{\J}^{{#1}#3}_{{#2}}}
\def\bpd#1#2#3{\bar{\J}^{{#1}}_{{#2}}{}_#3}
\def\viel#1#2{e_#1{}^#2} 
\def\gr#1#2#3{\C^{#1}_{#2{#3}}} 
\def\bgr#1#2#3{\bar{\C}^{#1}_{#2{#3}}}
 
\def\beq{\begin{equation}}
\def\eeq{\end{equation}}
\def\beqx{\begin{displaymath}} 
\def\eeqx{\end{displaymath}}
\def\beql{\arraycolsep .1em \begin{eqnarray}}
\def\eeql{\end{eqnarray}}
\def\zeile{\nonumber\\[.5ex] }
\def\gl#1{(\ref{#1})}
\def\sect#1{\setcounter{equation}{0}\section{#1}}

\begin{document}


\begin{titlepage}
\noindent
December, 1996 \hfill IASSNS-HEP-96/129 \\
\phantom{X}\hfill ITP--UH--27/96 \\
\phantom{X}\hfill hep-th/9612218

\vskip 1.0cm

\begin{center}

{\Large\bf Path-Integral Quantization of the (2,2) String~$^*$}\\

\vskip 1.5cm

{\large Jan Bischoff ${}^2$ \ and \ Olaf Lechtenfeld ${}^{1,2}$}

\vskip 1.0cm

{\it ${}^1$ School of Natural Sciences}\\
{\it Institute for Advanced Study}\\
{\it Olden Lane, Princeton, NJ 08540, U.S.A.}\\

\vskip 0.3cm
and
\vskip 0.3cm

{\it ${}^2$ Institut f\"ur Theoretische Physik}\\ 
{\it Universit\"at Hannover}\\
{\it Appelstra\ss{}e 2, 30167 Hannover, Germany}\\
{http://www.itp.uni-hannover.de/\~{}lechtenf/}\\

\vskip 2cm
\textwidth 6truein
{\bf Abstract}
\end{center}

\begin{quote}
\hspace{\parindent}
{}\ \ \
A complete treatment of the (2,2) NSR string in flat (2+2) dimensional
space-time is given, from the formal path integral over $N{=}2$ super
Riemann surfaces to the computational recipe for amplitudes at any
loop or gauge instanton number. We perform in detail the superconformal
gauge fixing, discuss the spectral flow, and analyze the supermoduli space
with emphasis on the gauge moduli. Background gauge field configurations
in all instanton sectors are constructed. We develop chiral bosonization
on punctured higher-genus surfaces in the presence of gauge moduli and
instantons. The BRST cohomology is recapitulated, with a new space-time
interpretation for picture-changing. We point out two ways of combining
left- and right-movers, which lead to different three-point functions.
\end{quote}

\vfill

\textwidth 6.5truein
\hrule width 5.cm
\vskip.1in

{\small
\noindent ${}^*$
supported in part by `Deutsche Forschungsgemeinschaft'
and `Volkswagen-Stiftung'\\
\noindent ${}^2$
permanent address
}

\eject
\end{titlepage}
\newpage

\sect{Introduction}
String theories with more than one worldsheet supersymmetries
have been somewhat neglected in the past, since their critical
real space-time dimension of $4{+}0$ or $2{+}2$ does not promise
interesting phenomenology (for a review before 1993, see \cite{marcus}).
This has changed recently, after connections to maximally 
helicity-violating QCD amplitudes \cite{bardeen,siegel} 
and to F theory \cite{vafaF} were suggested. 
Since massless kinematics require the choice of (2,2) signature, 
our work deals with this case exclusively.

It has been known for some time \cite{ooguri} that open and closed critical 
$N{=}2$ strings each contain each a single massless scalar excitation,
which describes self-dual Yang-Mills and self-dual gravity, respectively,
in four dimensions.
The string reproduces the known fact that there is almost no scattering
in these self-dual field theories, and its loop amplitudes should provide 
guidance for the construction of their proper quantum extensions.
Surprisingly, very little was known about $N{=}2$ strings beyond the
tree-level (see, however, \cite{bonini}), until Berkovits, Ooguri and
Vafa \cite{berkovits1,berkovits2,vafa} reformulated (2,2) strings 
in terms of (4,4) topological strings,\footnote{
see also \cite{gomis} for an earlier note}
providing an elegant technique for loop calculations.
In particular, they demonstrated how the choice of a complex structure 
in space-time breaks $N{=}4$ to $N{=}2$ and $SO(2,2)$ Lorentz symmetry 
to $U(1,1)$.

The traditional (2,2) NSR formulation should yield the same results
in a more elementary fashion. To support this claim,
the authors have, together with Ketov, developed the BRST quantization
of $N{=}2$ strings and its novelties compared to the $N{=}1$ 
case~\cite{BKL,KL,lechtalk}.
Recently, we have shown \cite{BL} that $SO(2,2)$ Lorentz invariance can be 
made manifest in the (2,2) formulation, when gauge instantons are taken
into account and because the string coupling rescales under Lorentz 
boosts~\cite{parkes}.

This work gives a concise and detailed account of the path-integral 
quantization for the (2,2) string in flat $\buC^{1,1}$ background,
using the NSR formulation. Such a treatment is available for the $N{=}1$
string \cite{gsw,dhoker}, but has been missing in the $N{=}2$ case.
Although it inevitably contains some material derived and applied 
in \cite{BKL,KL,lechtalk}, much of it (and certainly most details)
cannot be found in the literature, to our knowledge.
It is our aim to present a self-consistent basis for subsequent computations
of $N{=}2$ string amplitudes, and to provide the reader with the
necessary tools for such an undertaking.

The paper is organized as follows.
After introducing the classical (2,2) string and its symmetries in
Section 2, the gauge-fixing procedure in the path integral is worked
out extensively in Section 3. As an interlude, Section 4 covers
the superconformal constraint algebra, an operator realization of its
spectral flow automorphism, and the BRST symmetry.
Section 5 deals with the $N{=}2$ supermoduli space, focusing on the
$U(1)$ gauge moduli. In addition, two background gauge field 
configurations representing topologically non-trivial gauge bundles 
(nonzero Chern number or instanton sector) are constructed.
In Section 6, chiral bosonization in the presence of the gauge moduli
is developed, including the evaluation of the sum over bosonic soliton
sectors. An instanton-creation operator generalizing spectral flow 
emerges as a by-product. Physical states and vertex operators are the
subject of Section 7, which recaps the BRST cohomology 
and its surprises~\cite{bien,BKL,pope,BL}.
The conclusions, Section 8, compare two ways of joining left- and
right-movers on the example of the (full) tree-level 3-point function.
Three Appendices contain our conventions and give ample details 
on the spectral-flow operator and on punctured Riemann surfaces.

\sect{The classical (2,2) String}
The critical $N{=}2$ string lives in $(2,2)$ real or $(1,1)$ complex 
space-time dimensions. Let the bosonic string coordinate be denoted by
\beq
\zu{\pm}{\m}\ =\ \zu{1}{\m}\pm i\zu{2}{\m}, \qquad \m=\pm
\eeq
where the first superscript $\pm$ refers to complex conjugation 
(for real components $1$ and $2$) and 
the second superscript $\m$ refers to $(1,1)$ light-cone coordinates. 

The scalar product of two $(2,2)$ vectors $k,p$ can be expressed in this 
notation by
\beql
k\cdot p\ &=&\ \fr{1}{2}(k^{+}\cdot p^{-}+k^{-}\cdot p^{+})
\zeile   &=&\ \fr{1}{4}(k^{++}p^{--}+k^{+-}p^{-+}+k^{-+}p^{+-}+k^{--}p^{++}).
\eeql
The dot (on the r.h.s.) is understood as the $(1,1)$ scalar product whenever 
the factors carry one $\pm$ index.
We also introduce
\beql
k^+\wedge p^+\ &=&\ k^{++}p^{+-} - k^{+-}p^{++} \zeile
k^-\wedge p^-\ &=&\ k^{-+}p^{--} - k^{--}p^{-+}
\eeql
The same notation applies for the supersymmetric partner $\J$ of $Z$ 
which consists of a complex combination of two Majorana spinors,
\beq
\pu{\pm}{}{\m}\ =\ \pu{}{}{1\m}\pm i\pu{}{}{2\m},
\eeq
whose components are labeled by $\ua\da$-arrows,
\beq
\pu{\pm}{}{\m}\ =\ \left(\begin{array}{c}
                       \pu{\pm}{\ua}{\m} \\ \pu{\pm}{\da}{\m}
                       \end{array}\right).
\eeq
To construct the $N{=}2$ string action one has to couple the $N{=}2$ matter 
to the minimal $N{=}2$ supergravity multiplet living on the two-dimensional 
worldsheet. This supergravity multiplet contains 
the zweibein $\viel{m}{a}$, the gravitini $\gr{\pm}{m}{}$ and a 
$U(1)$ gauge field $A_{m}$. 

The $N{=}2$ supersymmetric string action in a flat space-time background
is given by \cite{brink} 
\beq
S_{m}\ =\ -\fr{1}{4\p}\int_{\cm} d^2 z\sqrt{g}\cl_m
\eeq
with matter Lagrangian~\footnote{
For the definitions of the covariant derivatives $D_m$ and $\cd_m$
consult Appendix A.}
\beql\label{lag1}
\cl_m\ &=&\ \fr{1}{2} g^{mn}\pa_m \zu{-}{\m} \pa_n \zd{+}{\m} 
         +\fr{i}{2} \bpu{-}{}{\m} \g^{m} \dvec{D}_m \pd{+}{}{\m} 
         +A_m \bpu{-}{}{\m} \g^{m} \pd{+}{}{\m} \zeile
      && +(\pa_m \zu{-}{\m}+\bpu{-}{}{\m} \gr{+}{m}{})
          \bgr{-}{n}{} \g^m \g^n \pd{+}{}{\m}
         +(\pa_m \zu{+}{\m}+\bgr{-}{m}{} \pu{+}{}{\m})
          \bpd{-}{}{\m} \g^n \g^m \gr{+}{n}{}.
\eeql
The complete set of local symmetries \cite{fradkin} acts infinitesimally
on the gravity and matter fields as
\beql\label{sym1}
\d\viel{m}{a}\ &=&\ \x^n \pa_n \viel{m}{a} + \viel{n}{a} \pa_m \x^n + 
                  l\ve^a{}_b \viel{m}{b}
             +2i(\bgr{-}{m}{} \g^a \ep{+}{} - \bep{-} \g^a\gr{+}{m}{}) + 
             \s\viel{m}{a},
\zeile
\d \gr{+}{m}{}\ &=&\ \x^n \pa_n \gr{+}{m}{} + \gr{+}{n}{}\pa_m \x^n  
               -\fr{1}{2}l\g_5 \gr{+}{m}{} + \cd_m\ep{+}{} \zeile
            && + \fr{1}{2}\s\gr{+}{m}{} +i\a\gr{+}{m}{} 
               + i\ha\g_5\gr{+}{m}{} 
               +\g_m \et{+},
\zeile
\d A_m\ &=&\ \x^n\pa_n A_m + A_n \pa_m \x^n+
           \ve^{nl}(\bep{-} \g_5 \g_m \cd \gr{+}{m}{}
                   -\cd_n \bgr{-}{l}{} \g_m \g_5\ep{+}{}) \zeile
       && +\pa_m\a + \ve_m{}^n\pa_n\ha + 
           \bgr{-}{n}{} \g_m \g^n \et{+} + \bet{-} \g^n \g_m \gr{+}{n}{},
\zeile
\d \zu{+}{\m}\ &=&\ \x^n \pa_n \zu{+}{\m} - 2\bep{-}\pu{+}{}{\m},
\zeile
\d \zu{-}{\m}\ &=&\ \x^n \pa_n \zu{-}{\m} - 2\bpu{-}{}{\m}\ep{+}{},
\zeile
\d \pu{+}{}{\m}\ &=&\ \x^n\pa_n\pu{+}{}{\m} - \fr{1}{2}l\g_5\pu{+}{}{\m} + 
                   i\g^n\ep{+}{}(\pa_n\zu{+}{\m} + 2(\bgr{-}{n}{}\pu{+}{}{\m}))
\zeile
                && - \fr{1}{2}\s\pu{+}{}{\m} + i\a\pu{+}{}{\m} 
                   + i\ha\g_5\pu{+}{}{\m},
\eeql
where the parameters and symmetries are related in the following way
\begin{center}
\begin{tabular}{lll}
$\x$ & $\ra$ & diffeomorphisms,\\
$\ep{\pm}{}$ & $\ra$ &$N=2$ supersymmetry transformations,\\
$l$  & $\ra$ & Lorentz transformation,\\
$\a$ & $\ra$ & $U_V(1)$ transformation,\\
$\ha$ & $\ra$ & $U_A(1)$ transformation,\\
$\s$ & $\ra$ & bosonic Weyl transformation,\\
$\h^{\pm}$ & $\ra$ & fermionic Weyl transformations.
\end{tabular}
\end{center}
In addition, \gl{lag1} is invariant under global $U(1,1)\times\bZ_2$ 
target space transformations which act on the $(1,1)$ complex indices 
of the matter fields $Z$ and $\J$ \cite{ooguri,BKL}.

For a planar worldsheet and topologically trivial $U_V(1)\times U_A(1)$
bundle these symmetries suffice to gauge away all gravitational 
degrees of freedom \cite{fradkin}. 
After switching to light-cone coordinates and performing the Wick rotation
to a Euclidean worldsheet, the resulting gauge-fixed Lagrangian 
\beq\label{freelag}
\cl_m^{fix}  = \pa_{z}\zu{-}{\m}\pa_{\bz}\zd{+}{\m} 
         +\pa_{\bz}\zu{-}{\m}\pa_{z}\zd{+}{\m}
         +\pu{-}{\ua}{\m}\dvec{\pa}_{z}\pd{+}{\ua}{\m} 
         +\pu{-}{\da}{\m}\dvec{\pa}_{\bz}\pd{+}{\da}{\m}
\eeq
has an enhanced target space symmetry, namely $SO(2,2)\supset U(1,1)$.
This gauge is globally appropriate only for tree-level 
and zero-instanton calculations of string amplitudes since beyond that
one has to cope with the $N{=}2$ moduli, 
the remnants of the gauge-fixing procedure.
Nevertheless, it can always be used locally, as long as one remembers to
compute the free field correlators in a non-trivial metric and gauge
background.

In the presence of vertex operators (which create infinitesimal holes
in the worldsheet) we need to specify the holonomies of the worldsheet
fermions around the punctures even on tree level.
The global subgroup of $U_V(1)\times U_A(1)$ acts as
\beql\label{monos}
\pu{\pm}{\da}{\m}\ &\longrightarrow\ e^{\pm 2\p i \r}\pu{\pm}{\da}{\m},
\qquad\qquad
\gr{\pm}{\da}{m}\ &\longrightarrow\ e^{\pm 2\p i \r}\gr{\pm}{\da}{m},
\zeile
\pu{\pm}{\ua}{\m}\ &\longrightarrow\ e^{\pm 2\p i \bar\r}\pu{\pm}{\ua}{\m},
\qquad\qquad
\gr{\pm}{\ua}{m}\ &\longrightarrow\ e^{\pm 2\p i \bar\r}\gr{\pm}{\ua}{m},
\eeql
so one can have a different constant pair $\r,\bar\r$ for each puncture
(and also for each standard homology cycle). Stated differently, the
worldsheet fermions are sections of a twisted holomorphic line bundle.
But there is more. We shall see that harmonic gauge transformations
serve to connect all different twists (for the same topology). 
The continuous transformation
\beq\label{flow}
(\r_\ell,\bar\r_\ell)\ \longrightarrow\ (\r'_\ell,\bar\r'_\ell)
\eeq
is called `spectral flow'.
Hence, no matter what choice of holonomies one starts with, 
the path integral is going to average over all of them.

\sect{Gauge Fixing and Path-Integral Quantization}
The appropriate description for higher-loop string interactions is given 
by the path integral over higher-genus worldsheet geometries. 
Since the 2d supergravity multiplet has no dynamical degrees of freedom 
it can be integrated out almost completely. 
In order to obtain a well-defined expression for the remaining 
path integral one has to split the gravitational fields in a 
gauge part and a non-gauge part, namely the $N{=}2$ moduli.
The gauge part can of course be expressed by the gauge parameters.
This change of variables in the path-integral measure will be accompanied
by a Jacobian which can be written as a path integral 
over ghost and antighost fields.
The integral over the gauge parameters is divergent and yields
the volume of the symmetry group which is factored out.

For the calculation of the Jacobian it is convenient to formulate the 
infinitesimal transformation laws of the supergravity multiplet in a 
covariant way.
Let us start with the zweibein,
\beq
h_b{}^a\ :=\ \viel{b}{m}\d\viel{m}{a}\ =\
           D_b\x^a -2i(\bep{-} \g^a \gr{+}{b}{}-\bgr{-}{b}{}\g^a\ep{+}{})
           -(l+\o_c\x^c)\e_b{}^a+\s\d_b^a -I^a\ve_{bc}\x^c,
\eeq
where $I^a = i\ve^{bc}\bgr{-}{b}{}\g^a\gr{+}{c}{}$.
Let us next consider the transformations for the gravitini 
and the $U(1)$ gauge field lifted to the tangent space:
\beql
\d\gr{\pm}{a}{}\ &=&\ \d(\viel{a}{m}\gr{\pm}{m}{})\ 
                 =\ -h_a{}^b\gr{\pm}{b}{}+\viel{a}{m}\d\gr{\pm}{m}{}
\zeile
&=&\ -h_a{}^b\gr{\pm}{b}{}+
    D_a\x^b\gr{\pm}{b}{}+\x^b D_b\gr{\pm}{a}{}+\cd_a \ep{\pm}{}
    \pm i\a\gr{\pm}{a}{}\mp i\g_5\ha\gr{\pm}{a}{},
\zeile
          &&-\fr{1}{2}(l+\o_c\x^c)\g_5\gr{\pm}{a}{}
            +\fr{1}{2}\s\gr{\pm}{a}{}+\g_a\et{\pm}{}
            -\gr{\pm}{c}{}\ve_{ab}\x^b I^c,
\zeile
\d A_a\ &=&\ \d(\viel{a}{m}A_{m})\ =\ -h_a{}^b A_b + \viel{a}{m}\d A_{m}
\zeile
&=&\ -h_a{}^b A_b+D_a\x^b A_b+\x^b D_b A_a 
    +\ve^{bc}(\bep{-}{}\g_5\g_a\cd_b\gr{+}{c}{}
             -\cd_b\bgr{-}{c}{}\g_a\g_5\ep{+}{})
\zeile
&&-I^d(\bep{-}{}\g_5\g_a\gr{+}{d}{}-\bgr{-}{d}{}\g_a\g_5\ep{+}{})
\zeile
&&+\pa_a\a+\ve_a{}^b\pa_b\ha+\bgr{-}{b}{}\g_a\g^b\et{+}{} 
    +\bet{-}{}\g^b\g_a\gr{+}{b}{}.
\eeql
To separate the ranges of the individual transformations it is necessary to 
redefine some of their parameters,
\beql
\ts\ &=&\ \s+\fr{1}{2}D_a\x^a, \zeile
\tl\ &=&\ l+\o_c\x^c+\fr{1}{2}(D_{\bz}\x^{\bz}-D_z\x^z), \zeile
\th^{\pm}\ &=&\ \h^{\pm}+\fr{1}{2}\g^a\cd_a\e^{\pm}.
\eeql
The first two redefinitions separate the diffeomorphisms,
Weyl and Lorentz transformations, 
whereas the third redefinition separates fermionic Weyl and 
supersymmetry transformations.
We can now eliminate the components $h_z{}^z$ and $h_{\bz}{}^{\bz}$ 
by combinations of Lorentz ($l$) and Weyl ($\s$) transformations and thus 
fix these two symmetries.

Using the inhomogenous fermionic Weyl transformations ($\h^{\pm}_{\ua\da}$), 
it is also possible to gauge away the components $\gr{\pm}{\bz}{\da}$ and 
$\gr{\pm}{z}{\ua}$ of the gravitini. 
Thus we will only have to introduce ghosts for the diffeomorphisms, 
supersymmetry and the $U_{V}(1), U_A(1)$ transformations. 
In order to separate holomorphically the range of the remaining unfixed 
symmetries (see \cite{atick}), 
one has to redefine the supersymmetry and $U(1)$ gauge parameters as follows, 
\beql
\tep{\pm}{\ua}\ &=&\ \ep{\pm}{\ua}+\x^{\bz}\gr{\pm}{\bz}{\ua},
\zeile
\tep{\pm}{\da}\ &=&\ \ep{\pm}{\da}+\x^{z}\gr{\pm}{z}{\da},
\zeile
\ta-\tha\ &=&\ \a-\ha+\x^{\bz}A_{\bz},
\zeile
\ta+\tha\ &=&\ \a+\ha+\x^{z}A_{z}.
\eeql
The final result for the covariant transformations of the zweibein,
the gravitini, and the gauge connection is given by
\beql\label{varAgr}
h_{\bz}{}^{z}\ &=&\ D_{\bz}\x^z 
- 4(\tep{-}{\ua}\gr{+}{\bz}{\ua}-\gr{-}{\bz}{\ua}\tep{+}{\ua}),
\zeile
h_{z}{}^{\bz}\ &=&\ D_z\x^{\bz} 
+ 4(\tep{-}{\da}\gr{+}{z}{\da}-\gr{-}{z}{\da}\tep{+}{\da}),
\zeile
\d\gr{\pm}{\bz}{\ua}\ &=&\ \x^z D_z\gr{\pm}{\bz}{\ua}
                   -\fr{1}{2}D_z\x^z\gr{\pm}{\bz}{\ua} 
                   +\cd_{\bz}\tep{\pm}{\ua} \pm i(\ta-\tha)\gr{\pm}{\bz}{\ua},
\zeile
\d\gr{\pm}{z}{\da}\ &=&\ \x^{\bz} D_{\bz}\gr{\pm}{z}{\da}
                   -\fr{1}{2}D_{\bz}\x^{\bz}\gr{\pm}{z}{\da}
                   +\cd_{z}\tep{\pm}{\da} \pm i(\ta+\tha)\gr{\pm}{z}{\da},
\zeile
\d A_{\bz}\ &=&\ \x^z D_z A_{\bz}-4iD_z\x^{\bz}\gr{-}{\bz}{\ua}\gr{+}{\bz}{\ua}
\zeile
        &&+2i[D_z\gr{+}{\bz}{\ua}\tep{-}{\ua}
             -D_z\gr{-}{\bz}{\ua}\tep{+}{\ua}
             +\gr{-}{\bz}{\ua}D_z\tep{+}{\ua}
             -\gr{+}{\bz}{\ua}D_z\tep{-}{\ua}]
          +\pa_{\bz}(\ta-\tha),
\zeile
\d A_{z}\ &=&\ \x^{\bz}D_{\bz}A_z-4iD_{\bz}\x^z\gr{-}{z}{\da}\gr{+}{z}{\da}
\zeile
        &&+2i[D_{\bz}\gr{+}{z}{\da}\tep{-}{\da}
             -D_{\bz}\gr{-}{z}{\da}\tep{+}{\da}
             +\gr{-}{z}{\da}D_{\bz}\tep{+}{\da}
             -\gr{+}{z}{\da}D_{\bz}\tep{-}{\da}]
          +\pa_{z}(\ta+\tha).
\eeql
It factorizes holomorphically.\footnote{
The anholomorphic piece of $\d A_{\bz}$, 
containing $\gr{-}{\bz}{\ua}\gr{+}{\bz}{\ua}$, disappears when choosing 
delta-function support for the gravitini (see Section 5.)}
In particular, the linear combinations $U_V(1)\pm U_A(1)$ act
independently on left- and right-movers.
We are now able to construct the ghost action directly from the 
transformation laws \gl{varAgr}.

For the remainder of this section, 
we will concentrate only on the holomorphic half of \gl{varAgr},
i.e. its first, third, and fifth equation.
An arbitrary variation of the gravitational multiplet can be achieved 
by a gauge transformation \gl{varAgr} plus an infinitesimal shift in the 
$N{=}2$ moduli space,
\beq\label{allgvar}
\left(\begin{array}{c}
      \tilde{h}_{\bz}{}^{z}\\ \tilde{\d}\gr{\pm}{\bz}{\ua}\\ \tilde{\d} A_{\bz}
      \end{array}\right)\
=\ \cp\left(\begin{array}{c}
             \x^{z}\\ \tep{\pm}{\ua}\\ \ta-\tha
             \end{array}\right) 
+ \left(\begin{array}{ccc}
        \m_{i\bz}^{z}&0&0\\0&\n^{\pm}_{j}&0\\0&0&\o_{k\bz}
        \end{array}\right)
  \left(\begin{array}{c}
        t^{i}\\ \z^{j\pm}_{\bz\ua}\\C^{k}
        \end{array}\right)\quad.
\eeq
The operator $\cp$ in \gl{allgvar} is read off the holomorphic half
of \gl{varAgr} as
\beq\label{Pop}
\cp\ =\ \left(\begin{array}{cccc}
		D_{\bz} & 4\gr{-}{\bz}{\ua} & 4\gr{+}{\bz}{\ua} & 0 \\ 
		D_z\gr{+}{\bz}{\ua}-\fr{1}{2}\gr{+}{\bz}{\ua}D_z &
			\cd_{\bz} & 0 & +i\gr{+}{\bz}{\ua} \\
		D_z\gr{-}{\bz}{\ua}-\fr{1}{2}\gr{-}{\bz}{\ua}D_z &
			0 & \cd_{\bz} & -i\gr{-}{\bz}{\ua} \\
		D_z A_{\bz} & 
		\quad -2i D_z\gr{-}{\bz}{\ua}+2i\gr{-}{\bz}{\ua}D_z \quad &
		\quad  2i D_z\gr{+}{\bz}{\ua}-2i\gr{+}{\bz}{\ua}D_z \quad &
			\pa_{\bz}
		\end{array}\right)
\eeq
whereas
$\m_{i\bz}^{z}, \n^{\pm}_{j}$, $\o_{k\bz}$ are tangent vectors of the 
$N{=}2$ moduli space with coordinates $t^{i}, \z^{j\pm}_{\bz\ua}, C^{k}$. 

The relevant Jacobian is
\beql\label{sdet}
\sdet'\cp\ &=&\ \int [dcdb][d\g^{\pm} d\b^{\mp}][d\tc d\tb]\;
  e^{-S_{gh}[c,b,\g^{\pm},\b^{\mp},\tc,\tb]} \zeile
&&\times \prod_{i}\VEV{\m_{i}|b}\prod_{k}\langle\o_{k}|\tb\rangle
  \prod_{j}\d(\langle\n^{+}_{j}|\b^{-}\rangle)
  \prod_{j}\d(\langle\n^{-}_{j}|\b^{+}\rangle) \quad,
\eeql
where the ghost action 
\beq
S_{gh}\ =\ \fr{1}{\p}\int d^{2}z\sqrt{g}\cl_{gh}
\eeq
is obtained by sandwiching $\cp$ between
the ghost triple $(c,\g^\pm,\tc)$
and the antighost triple $(b,\b^\mp,\tb)$
dual to $(h_{\bz}{}^{z},\d\gr{\pm}{\bz}{\ua},\d A_{\bz})$.
One finds
\beql
\cl_{gh}\ &=&\ bD_{\bz}c+\b^{-}D_{\bz}\ga{+}+\b^{+}D_{\bz}\ga{-}
             +\tb\pa_{\bz}\tc
\zeile
         &&+\gr{-}{\bz}{\ua}[-4b\ga{+}+D_z(c\b^{+})+\fr{1}{2}\b^{+}D_zc 
           -\b^{+}\tc-2D_z\tb\ga{+}-4\tb D_z\ga{+}]
\zeile
         &&+[+4b\ga{-}-D_z(c\b^{-})-\fr{1}{2}\b^{-}D_zc 
           -\b^{-}\tc-2D_z\tb\ga{-}-4\tb D_z\ga{-}]\gr{+}{\bz}{\ua}
\zeile
         &&-iA_{\bz}[\b^{-}\ga{+}-\b^{+}\ga{-}-D_z(c\tb)]
           -(4\tb\gr{-}{\bz}{\ua}\gr{+}{\bz}{\ua}D_z\bar{c})+\hbox{tot. der.}
\eeql
The prime on the super determinant in \gl{sdet} indicates that
the kernel of the adjoint operator $\cp\dg$ is to be projected out,
since it consists of the modes outside the image of $\cp$.
The projection is achieved by the antighost insertions in the
integrand of \gl{sdet}.
The kernel of $\cp$ itself spans the isometries of the super-worldsheet 
and must be compensated for by vertex operator insertions.
For a locally flat gauge $(A=\c=0, e=1)$ one obtains
\beq
\cl_{gh}^{fix}\ =\  
b\pa_{\bz}c+\b^{-}\pa_{\bz}\ga{+}+\b^{+}\pa_{\bz}\ga{-} +\tb\pa_{\bz}\tc.
\eeq

Collecting all results we get the contribution of a fixed topology
to the partition function of the $N{=}2$ string,
\beql\label{zsum}
\cz\ &=&\ \int\frac{[d\viel{m}{a}][d\gr{\pm}{m}{}][dA_{m}]}
{\hbox{symmetry volume}}
	[dZ][d\J]\; e^{-S_{m}[e,\c^\pm,A;Z,\J]}\zeile
&=&\int\Bigl| dtd\z^{\pm}dC \Bigr|^2 
	\int\Bigl| [dcdb][d\g^\pm d\b^\mp][d\tc d\tb] \Bigr|^2
	[dZ][d\J]\; e^{-S_m^{fix}-S_{gh}^{fix}-\bar S_{gh}^{fix}} \zeile
&&\times \Bigl| \prod_{i}\VEV{\m_{i}|b}\prod_{k}\langle\o_{k}|\tb\rangle
  \prod_{j}\d(\langle\n^{+}_{j}|\b^{-}\rangle)
  \prod_{j}\d(\langle\n^{-}_{j}|\b^{+}\rangle) \Bigr|^2 .
\eeql

\sect{Superconformal Algebra, Spectral Flow and BRST}
Having derived the complete (matter plus ghost) action away from a 
locally flat gauge, we are able to calculate the
ghost-extended symmetry currents that form the anomaly-free 
$N{=}2$ superconformal constraint algebra, 
\beq
\hT_{ab}\ =\ -\fr{1}{\sqrt{g}}\h_{ac}\viel{m}{c}
            \parvar{\sqrt{g}\cl}{\viel{m}{b}}, \qquad
\hGe{\pm}_{a\ua\da}\ =\  
          -\fr{1}{4}\parvar{\cl}{\gr{\mp}{a}{\da\ua}}, \qquad
\hJ_{a}\ =\ -\fr{i}{4}\parvar{\cl}{A_{a}}.
\eeq  
In a locally flat gauge these currents factorize holomorphically, with
\beql\label{symcur}
\hT_{zz}\ &=&\ -\fr{1}{2}\pa_{z}\zu{-}{\m}\pa_{z}\zd{+}{\m} 
             -\fr{1}{4}\pa_{z}\pu{-}{\da}{\m}\pd{+}{\da}{\m} 
             -\fr{1}{4}\pa_{z}\pu{+}{\da}{\m}\pd{-}{\da}{\m} \zeile
           &&+ 2\pa cb+c\pa b +\pa\tc\tb 
             -\fr{3}{2}(\pa\ga{-}\b^{+}+\pa\ga{+}\b^{-})
             -\fr{1}{2}(\ga{-}\pa\b^{+}+\ga{+}\pa\b^{-}), 
\zeile
\hGe{\pm}_{z\da}\ &=&\ \pa_{z}\zu{\mp}{\m}\pd{\pm}{\da}{\m} \zeile
                 &&-4\ga{\pm}b\mp 4\pa\ga{\pm}\tb\mp 2\ga{\pm}\pa\tb
                   +\fr{3}{2}\pa c\b^{\pm}+c\pa\b^{\pm}\mp\tc\b^{\pm},
\zeile
\hJ_{z}\ &=&\ -\fr{1}{2}\pu{-}{\da}{\m}\pd{+}{\da}{\m} \zeile
          &&+\pa(c\tb)+(\ga{+}\b^{-}-\ga{-}\b^{+}),
\eeql
decomposing into matter and ghost parts, e.g. $\hT=T_m+T_{gh}$.
Canonical quantization of the $N{=}2$ string is achieved by imposing the 
usual operator product expansions for the matter fields,
\beq
\zu{\pm}{\m}(z)\zu{\mp}{\n}(w)\ \sim\ -2\h^{\m\n}\log(z-w),
\qquad
\pu{\pm}{\da}{\m}(z)\pu{\mp}{\da}{\n}(w)\ \sim\ -2\h^{\m\n}(z-w)^{-1},
\eeq
and for the ghost fields,
\beq
\be{\pm}(z)\ga{\mp}(w)\ \sim\ \frac{-1}{z-w}, \qquad
b(z)c(w)\ \sim\ \frac{1}{z-w}, \qquad
\tb(z)\tc(w)\ \sim\ \frac{1}{z-w}.
\eeq
The operator product algebra is then given by
\beql\label{sca}
\hT(z)\ \hT(w)\ &\sim&\ \frac{c}{2}(z-w)^{-4} 
	+ 2(z-w)^{-2}\hT(w) + (z-w)^{-1}\pa\hT(w),
\zeile
\hT(z)\ \hGe{\pm}(w)\ &\sim&\ \frac{3}{2}(z-w)^{-2}\hGe{\pm}(w)
        + (z-w)^{-1}\pa\hGe{\pm}(w),
\zeile
\hT(z)\ \hJ(w)\ &\sim&\ (z-w)^{-2}\hJ(w) + (z-w)^{-1}\pa\hJ(w),
\zeile
\hGe{+}(z)\ \hGe{-}(w)\ &\sim&\ \frac{4c}{3}(z-w)^{-3} 
	- 4(z-w)^{-2}\hJ(w) + 4(z-w)^{-1}(\hT(w)-\fr{1}{2}\pa\hJ(w)),
\zeile
\hGe{\pm}(z)\ \hGe{\pm}(w)\ &\sim&\ 0,
\zeile
\hJ(z)\ \hGe{\pm}(w)\ &\sim&\ \mp (z-w)^{-1}\hGe{\pm}(w),
\zeile
\hJ(z)\ \hJ(w)\ &\sim&\ \frac{c}{3}(z-w)^{-2}.
\eeql
with central charge $c=c_m+c_{gh}=6-6=0$. 

The superconformal algebra \gl{sca} possesses a continuous automorphism
termed spectral flow \cite{lerche}.  It can be generated by
an operator which is constructed out of the $U(1)$ gauge current $J$ 
as~\footnote{ For a derivation see Appendix B.}
\beq\label{sfodef}
\SFO(\Q)\ =\ \exp\Bigl[\Q\oint\frac{dz}{2\pi i}\log(z)J(z)\Bigr],
\qquad \SFO(\Q)^{-1}=\SFO(-\Q).
\eeq
and acts on the currents as follows:
\beql
T(z)\quad&\ra&\quad\SFO(\Q)T(z)\SFO(\Q)^{-1}\
                =\ T(z)+\frac{\Q}{z}J(z)+\frac{c}{6}\frac{\Q^{2}}{z^{2}},
\zeile
G^{\pm}(z)\quad&\ra&\quad\SFO(\Q)G^{\pm}(z)\SFO(\Q)^{-1}\
                            =\ z^{\mp\Q}G^{\pm}(z),
\zeile
J(z)\quad&\ra&\quad\SFO(\Q)J(z)\SFO(\Q)^{-1}\ =\ J(z)+\frac{c}{3}\frac{\Q}{z}.
\eeql
The definition of $\SFO$ should be understood as a successive application 
of operator products, with the integrals of $J$ being rather formal. 
For example, 
\beql
T(w)\ &\ra&\ 
T(w)+\Q\oint\frac{dz}{2\pi i}\log(z)J(z)T(w)
	+\frac{\Q^2}{2}\oint\frac{dz'}{2\pi i}\log(z')J(z')
       		         \oint\frac{dz}{2\pi i}\log(z)J(z)T(w)
\zeile
&=&\ T(w)+\Q\oint\frac{dz}{2\pi i}\log(z)\frac{J(w)}{(z-w)^{2}}
	+\frac{\Q^2}{2}\oint\frac{dz'}{2\pi i}\log(z')J(z')
                         \oint\frac{dz}{2\pi i}\log(z)\frac{J(w)}{(z-w)^{2}}
\zeile
&=&\ T(w)+\frac{\Q}{w}J(w)
	+\frac{c\Q^2}{6}\oint\frac{dz'}{2\pi i}\log(z')J(z')\frac{J(w)}{w}
\zeile
&=&\ T(w)+\frac{\Q}{w}J(w)+\frac{c}{6}\frac{\Q^{2}}{w^2}.
\eeql
For the hatted currents one finds of course $c=0$.
A variant of this $\SFO$ will arise automatically in the evaluation 
of the path integral for the $N{=}2$ string.
Similar considerations apply independently for the anti-holomorphic copy
of the superconformal algebra.

The same ghost-extended symmetry currents \gl{symcur} may be calculated by 
using the canonical quantization procedure and BRST methods.
The resulting (holomorphic part of the) BRST current is given by
\beql
J_{BRST}\ &=&\ cT+\ga{+}\Ge{-}+\ga{-}\Ge{+}+\tc J +c\pa cb+c\pa\tc\tb
\zeile
         &&-4\ga{+}\ga{-}b+2\pa\ga{-}\ga{+}\tb-2\pa\ga{+}\ga{-}\tb
\zeile
         &&+\fr{3}{4}\pa c(\ga{+}\b^{-}+\ga{-}\b^{+})
           -\fr{3}{4}c(\pa\ga{+}\b^{-}+\pa\ga{-}\b^{+})
\zeile
         &&+\fr{1}{4}c(\ga{+}\pa\b^{-}+\ga{-}\pa\b^{+})
           +\tc(\ga{+}\b^{-}-\pa\ga{-}\b^{+}),
\eeql
and the BRST charge is defined by
\beq
Q\ =\ \oint \frac{dz}{2\p i} J_{BRST}.
\eeq
The ghost-extended currents then obey
\beq\label{hutstrom}
\hT(w)\ =\ \{Q,b(w)\}, \qquad
\hGe{\pm}(w)\ =\ [Q,\b^{\pm}(w)], \qquad
\hJ(w)\ =\ \{Q,\tb(w)\}.
\eeq
For completeness we list the BRST transformations 
of all matter and ghost fields in the superconformal gauge,
\beql
[Q,\zu{\pm}{\m}]\ &=&\ c\pa\zu{\pm}{\m}-2\ga{\mp}\pu{\pm}{}{\m},
\zeile
\{Q,\pu{\pm}{}{\m}\}\ &=&\ c\pa\pu{\pm}{}{\m}+\fr{1}{2}\pa c\pu{\pm}{}{\m}
                      -2\ga{\pm}\pa\zu{\pm}{\m}\mp\tc\pu{\pm}{}{\m},
\zeile
\{Q,c\}\ &=&\ c\pa c-4\ga{-}\ga{+},
\zeile
[Q,\ga{\pm}]\ &=&\ c\pa\ga{\pm}-\fr{1}{2}\pa c\ga{\pm}\mp\tc\ga{\pm},
\zeile
\{Q,\tc\}\ &=&\ c\pa\tc-2\ga{-}\pa\ga{+}+2\ga{+}\pa\ga{-}.
\eeql
The invariant (Neveu-Schwarz) vacuum is defined by the vanishing of
all $N{=}2$ currents at the origin $z=0$, which translates to
\beq
\hat{L}_{n\ge-1}\ket{0}\ = 0, \qquad
\hat{G}_{r\ge-1/2}\ket{0}\ = 0, \qquad
\hat{J}_{n\ge0}\ket{0}\ = 0.
\eeq
The subalgebra of \gl{sca} which leaves the vacuum invariant
is the $SL(2,\bR)\times U(1)$ generated by 
$\hat{L}_{\pm1},\hat{L}_{0}$ and $\hat{J_{0}}$. 
Again, one ultimately has to adjoin the antiholomorphic sector as well. 

\sect{The (2,2) Moduli Space and Instanton Background}
The space of moduli is defined as the complement to the range of 
gauge transformations. Since a gauge transformation is usually mediated 
by a differential operator $D$ and a transformation parameter $\x$, 
one arrives at
\beq\label{mod}
m\in \{\hbox{moduli}\} \qquad\Leftrightarrow\qquad 
0=\VEV{m|D\x}=\langle{D\dg m|\x}\rangle \quad\forall \x \qquad
\Leftrightarrow \qquad m\in\ker(D^{\dg}).
\eeq
The most powerful tool for the investigation of the $N{=}2$ moduli 
space is the Atiyah-Singer index theorem. With this theorem it is 
possible to determine the dimensions of the kernels of 
differential operators like in \gl{mod}.
For a Lorentz and gauge covariant derivative 
$\cd^{n,q}_{\bz}=\pa_{\bz}-n\o_{\bz}+2iqA_{\bz}$ 
on the space of $(-n)$-forms
over a compact Riemann surface with $p$ punctures
one obtains the relation ($n\in\fr12\bZ,\ q\in\bZ$)
\beq\label{index}
\dim\ker(\cd_{\bz}^{n,q}\dg)-\dim\ker(\cd_{\bz}^{n,q})\ =\ (2n+1)(g-1)+2qc+p,
\eeq
where $g\in\bZ_+$ is the genus of the Riemann surface
and $c\in\bZ$ is the first Chern number (instanton number)
of the $U(1)$-bundle 
(not to be confused with the diffeomorphism ghost or the central charge).
Generically, one of the two kernels is empty, depending on the
sign of the r.h.s. of \gl{index}.
In case $p=0$ and $q=0$, an additional zero mode will occur simultaneously
in both kernels for ($n=0, g\ge1$) or for (any $n, g=1$, odd spin structure).

For the $N{=}2$ string the appropriate weights $n$ and charges $q$ 
should be taken from the diffeomorphisms $\x^{z}$, the susy transformations 
$\e^{\pm}$ and the gauge transformation $\a$. 
For a genus $g>1$ and $p$ punctures one gets
\begin{center}
\begin{tabular}{lll}
$n=1, q=0$: & $3g-3+p$ & metric moduli $\m$\\
$n=\fr12, q=\pm\fr12$:\qquad{} & $2g-2\pm c+p$ & fermionic moduli $\n^\pm$\\
$n=0, q=0$: & $g-1+p+\d_{p,0}$\quad{} & gauge moduli $\o$
\end{tabular}
\end{center}
For a Chern number $\abs{c}>2(g-1)+p$, one index of \gl{index} 
turns negative, due to additional zero modes of $\g^+$ or $\g^-$.
In the path integral, these additional zero modes cannot be compensated for
and let the path integral vanish \cite{berkovits1} outside the range of
\beq\label{restr}
\abs{c}\ \le\ 2(g-1)+p.
\eeq
The dimensionalities given above are complex, but moduli space has
a complex structure, allowing one to consider left- and right-moving
moduli independently.

Let us investigate the properties of the fermionic moduli 
in more detail. As was shown in Ref.\cite{atick} it is possible 
to choose a special basis in fermionic moduli space which has a 
pointlike support. In that way we are able to parametrize the 
(variation of the) gauge-fixed gravitini as
\beq\label{fermod}
\d\gr{\pm}{\bz}{\ua}\ =\ \gr{\pm}{\bz}{\ua}\ =\
\sum_{j=1}^{2g-2\pm c+p}\z^{\pm j}_{\bz\ua}\,\d^2(z{-}z^{\pm}_j),
\qquad {\rm i.e.}\quad \n^{\pm j}\ =\ \d^2(z{-}z^{\pm}_j).
\eeq
With different points of support for $\gr{+}{\bz}{\ua}$ and 
$\gr{-}{\bz}{\ua}$, the annoying term quadratic in $\gr{\pm}{\bz}{\ua}$
vanishes from the ghost action. Furthermore, 
the expression \gl{fermod} can be inserted into the path integral \gl{zsum}. 
The finite-dimensional integral over the fermionic coordinates $\z^{\pm}$ 
is carried out rather easily \cite{verlinde2} and yields ($p=0$)
\beql\label{zgc}
\cz(g,c)\ &=&\ \int\Bigl| dtdC \Bigr|^2 
	\int\Bigl| [dcdb][d\g^\pm d\b^\mp][d\tc d\tb] \Bigr|^2 [dZ][d\J]\;
	e^{-S^{fix}[t,C;Z,\J,c,b,\ga{\pm},\b^{\mp},\tc,\tb]} \zeile
&&\times\Bigl| 
	\prod_{i=1}^{3g-3}\VEV{\m_i|b}\prod_{k=1}^{g}\langle\o_k|\tb\rangle 
	\prod_{j=1}^{2g-2+c}\!\d(\b^-(z^{+}_{j}))\hat{G}^{-}(z^{+}_{j})\!
	\prod_{j=1}^{2g-2-c}\!\d(\b^{+}(z^{-}_{j}))\hat{G}^{+}(z^{-}_{j})
	\Bigr|^2. 
\eeql
The combination $\d(\b)\hat{G}$ can 
be identified with the picture-changing operators \cite{FMS,BKL}
\beq
\PCO^{\pm}(z)\ =\ \{Q,H(\b^{\pm}(z))\} 
               =\ \d(\b^{\pm}(z))\{Q,\b^{\pm}(z)\} 
               =\ \d(\b^{\pm}(z))\hat{G}^{\pm}(z) \quad,
\eeq
where $H$ denotes the Heaviside step function. 
String scattering amplitudes ($p>0$) are obtained from \gl{zgc}
by shifting the number of moduli according to the table above
and by inserting a string of $p$ vertex operators in their
canonical representation (discussed in Section 7).
In any case, the fermionic moduli are responsible for the correct number of 
picture-changing operator insertions needed to balance the zero modes
of the $U(1)$ charged fields.

A new ingredient compared to the $N{=}1$ string is the gauged $U(1)$ 
symmetry and its moduli space which should be treated with care.
First of all, $U(1)$ bundles are topologically classified by the integral 
Chern number 
\beq\label{chern}
c\ =\ \fr{1}{2\p}\int F,\qquad\qquad F\ =\ dA \qquad\hbox{locally}
\eeq
which we have already encountered in the index theorem. 
Since the space of connections is an affine space, it is possible to 
split the $U(1)$ connection $A$ into a topologically trivial one, $A_{0}$ 
which is integrated over in the path integral, 
and a fixed background connection, $A_{c}$ which yields the Chern number,
\beq
A\ =\ A_{c} + A_0, \qquad\quad\hbox{with}\qquad
\int F_{c}\ =\ 2\p c \qquad\hbox{and}\qquad
\int F_{0}\ =\ 0.
\eeq
All continuous degrees of freedom are left in $A_{0}$ which,
as a globally defined one-form, may be Hodge-decomposed into 
exact, co-exact and harmonic parts, 
\beq\label{hodge}
A_0\ =\ d\l\ +\ i*d\m\ +\ h.
\eeq
A peculiarity of $N{=}2$ string theory is the fact that the
connection $A$ transforms under a $U_{V}(1)$ with parameter $\a$ 
as well as under a $U_{A}(1)$ with parameter $\hat{\a}$ (see \gl{sym1}),
\beql
A^{\a}\ &=&\ e^{-i\a}(A +id)e^{i\a}\ =\ A - d\a, 
\qquad\qquad F^{\a}\ =\ F,
\zeile
A^{\hat\a}\ &=&\ e^{-i\hat\a}(A -*d)e^{i\hat\a}\ =\ A - i*d\hat{\a} 
\qquad\quad F^{\hat\a}\ =\ F - *\D\hat\a.
\eeql
With these two transformations the exact and the co-exact parts in \gl{hodge}
can be completely eliminated so that only the harmonic contribution remains. 
In other words, one may choose a gauge ($\l=\f=0$) where
\beq
A_0\ =\ h \qquad\qquad\Longrightarrow\qquad\qquad F_0\ =\ 0\ =\ d*A_0,
\eeq
a flat connection in the Lorentz gauge.

On a compact punctured Riemann surface there exists a 
$2(g{+}p{-}1)$ dimensional basis 
of real harmonic one-forms $\a_i,\b_i$, $i=1,\ldots,g$, 
and $\g_\ell,\d_\ell$, $\ell=1,\ldots,p{-}1$,
dual to a homology basis of cycles $a_i,b_i$ and $c_\ell,d_\ell$.
Hence, an element of the $U(1)$ Teichm\"uller space may be written as
\beq 
h\ =\ 2\p( A^i\a_i + B^i\b_i + G^\ell\g_\ell + D^\ell\d_\ell )
\quad\hbox{with}\quad\vec{A},\vec{B}\in\bR^g
\quad\hbox{and}\quad\vec{G},\vec{D}\in\bR^{p-1}.
\eeq
Note that the $\ell$ sum runs only to $p{-}1$, 
because the sum of all puncture holonomies must vanish.
The coefficient vectors $\vec{A},\vec{B},\vec{G},\vec{D}$
can take any real values.
When all are integral, however, $g(z)=\exp\{i\int_p^z h\}$
is single-valued and generates a ``large'' gauge transformation
not connected to the identity.
Dividing out these ``modular'' gauge transformations compactifies
the $U(1)$ Teichm\"uller space $\bR^{2(g+p-1)}$ to the moduli space
$(\bR/\bZ)^{2(g+p-1)}$, since all coefficients become periodic on $[0,1)$.
It is convenient to combine the twists to $(g{+}p{-}1)$-dimensional vectors
$\tA=(\vec{A},\vec{G})$ and $\tB=(\vec{B},\vec{D})$ and
rewrite the flat connection in terms of a vector $\O$ 
of meromorphic one-forms $\o_{i}$ and $\o_{P_iP_{i+1}}$, 
\beql\label{flatA}
h\ &=&\ 2\p(\tB-\bT\tA)(T-\bT)^{-1}\O - 2\p(\tB-T\tA)(T-\bT)^{-1}\bO \zeile
  &=&\ -i\p \bar{C}(\Im T)^{-1}\O+i\p C(\Im T^{-1})\bO
\qquad\hbox{with}\quad C=\tB{-}T\tA\ \in\buC^{g+p-1}/\L,
\eeql
using the generalized period matrix $T$ and Jacobian lattice $\L$. 
As a sum of a meromorphic and an anti-meromorphic piece,
$h$ has simple poles at the punctures, with residues given by
$(G^{\ell}{-}G^{\ell-1},D^{\ell}{-}D^{\ell-1})$ for $\ell=1,\ldots p{-}1$ 
and $(G^p,D^p)=(-G^{p-1},-D^{p-1})$, making their sum vanish.
Appendix C contains further details.
Some care is required with the puncture (co)homologies.
The punctured surface may be regarded as a degeneration limit of
an unpunctured surface of genus $g{+}p{-}1$. In this limit, $p{-}1$ harmonic
one-forms $\d_\ell$ formally disappear since the $d_\ell$ cycles
blow up. The meromorphic one-forms built with them, however, remain finite.
Although their contributions to $h$ in \gl{flatA} vanish, as diagonal
elements of $\Im T$ diverge, their singular integrals along $d_\ell$
produce finite twists $D^\ell$.
Note that in the holomorphic basis the spectral flow acts on the 
gauge puncture moduli by simple shift, 
and holomorphic factorization is manifest.

Let us consider some possible background configurations $A_{c}$ explicitly.
For this purpose it will be sufficient to consider a compact Riemann
surface of genus $g$ without any punctures ($p=0$).
A non-vanishing Chern number $c$ means that we have 
to introduce more than one coordinate patch on the Riemann surface,
with non-trivial transition functions in their overlap.
To be more precise, only the $U_{V}(1)$-bundle may be non-trivial.
The $U_{A}(1)$ transition functions must stay trivial in order 
to have a well-defined Chern number. 
This is markedly different from \cite{berkovits2,vafa,BL} 
where formally {\it two\/} Chern numbers were introduced, 
one for the holomorphic and one for the anti-holomorphic half of the theory.
Such a prescription would in fact redefine the $(2,2)$ string,
in analogy to the GSO projection of the $(1,1)$ string,
where in the sum over spin structures one replaces the {\it common\/}
holonomies for left- and right-movers by {\it independent\/} ones.
Since spectral flow can be generalized to change the Chern number
(see below), it is natural to introduce left- and right-moving
Chern numbers, at the expense of giving up the local worldsheet description.
We shall see, however, that there is no need for a GSO projection in the
$N{=}2$ string, and one therefore has a choice here.
In the present work we will not follow \cite{berkovits2,vafa,BL},
but keep the geometrical path-integral definition. 

Our requirement for a trivial $U_{A}(1)$ bundle may be compared 
with the nature of diffeomorphisms and Weyl transformations. 
In order to have a well-defined genus, one has to 
require a globally defined Weyl parameter $\s$ so its contribution 
to the curvature scalar $\car$ is a total derivative.
The same applies for $c$ and $U_{A}(1)$,
\beql\label{analogy}
-8\p(g{-}1) = \int\!\sqrt{g}\car^\s = \int\!\sqrt{g}(\car + \D\s),
\qquad 2\p c = \int\!F^{\hat\a} = \int(F -*\D\ha).
\eeql
For that reason we only have to consider non-trivial $U_{V}(1)$ transition 
functions for the construction of some $A_{c}$. 

One way to cover a compact Riemann surface with coordinate 
patches is to cut it along a homology basis ${a_i,b_i}$,
demonstrated in Figure 1.
By this procedure one obtains a simply-connected region with a boundary 
as given in Figure 2.
\begin{figure}[t]
\begin{center}
\leavevmode\epsfxsize=10cm
\epsfbox{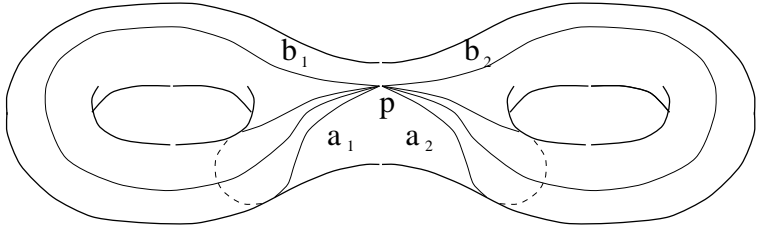}
\caption{Genus 2 Riemann surface.}
\end{center}
\end{figure}
\begin{figure}[t]
\begin{center}
\leavevmode\epsfxsize=6cm
\epsfbox{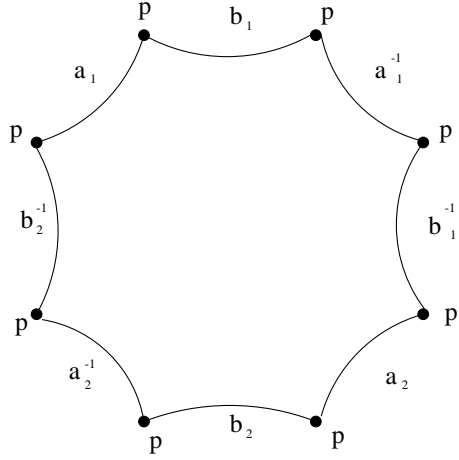}
\caption{Cut genus 2 Riemann surface.}
\end{center}
\end{figure}
The edges of this region, $a_i,a_i^{-1}$ respectively $b_i,b_i^{-1}$, 
have to be identified in order to reconstruct the original surface. 
On these edges one may install the non-trivial transition functions 
of the $U_V(1)$ bundle. 

For practical reasons we choose the Lorentz gauge for the connection,
\beq
d*A\ =\ \pa_z A_{\bz} + \pa_{\bz} A_z\ =\ 0.
\eeq
The possible transition functions $g=e^{i\a}$ are then restricted 
to have a harmonic exponent,
\beq
\pa_z\pa_{\bz}\a\ =\ 0 \qquad\Longleftrightarrow\qquad
\a(z,\bz)\ =\ \a(z) + \bar{\a}(\bz).
\eeq
The form of a finite $U_V(1)$ transition function $g$ on the homology cycles
$a_i,b_i$ can be built from the harmonic one-forms $\a_i,\b_i$,
\beql\label{transi}
g_{a_i}\ &=&\ \exp\{-2\p i\int_{p}^{z}
          \sum_{j=1}^{g}(\a_j u_{ji}^a+\b_j v_{ji}^a)\}, \zeile
g_{b_i}\ &=&\ \exp\{-2\p i\int_{p}^{z}
          \sum_{j=1}^{g}(\a_j u_{ij}^b+\b_j v_{ij}^b)\}, 
\eeql
where $g_{a_i},g_{b_i}$ live on the cycles $a_i,b_i$ which 
have to be crossed when walking around a cycle $b_i,a_i$.
If one requires a transition function of this type to be single-valued 
on the {\it whole\/} Riemann surface, one has to identify all coefficients 
$u,v$ on different cycles and take them to be integral, 
which yields just a ``large'' gauge transformation.

A simple choice of a connection and field strength with Chern number $c$ 
employs only the harmonic forms $\a_i,\b_i$, and is given by
\beq\label{anullhol}
A_{c}\ =\ \fr{\p ic}{2g}[\o\ (\Im\t)^{-1}\int_p^z\!\bo -
                         \bo\ (\Im\t)^{-1}\int_p^z\!\o]
\qquad\Longrightarrow\qquad
F_{c}\ =\ -\fr{\p ic}{g}\ \o\wedge(\Im\t)^{-1}\bo. 
\eeq
For the holonomies of $A_{c}$ we get
\beql\label{mono}
A_{c}(z+a_k)-A_{c}(z)\ &=&\ \fr{\p ic}{2g}(\o_i+\bo_i)(\Im\t)^{-1}_{ik}\ 
=\ ig_{b_k}^{-1}dg_{b_k},
\zeile
A_{c}(z+b_k)-A_{c}(z)\ &=&\ \fr{\p ic}{2g}(\o_i(\Im\t)^{-1}_{ij}\bt_{jk}- 
                                     \bo_i(\Im\t)^{-1}_{ij}\t_{jk})\
=\ ig_{a_k}^{-1}dg_{a_k}
\eeql
where $z+a_{k},b_{k}$ denotes a shift of $z$ along 
that homology cycle.
The parameters $u,v$ in \gl{transi} are then specified via
\beql
g_{b_k}\ &=&\ \exp\Bigl[\fr{\p c_1}{2g}\ e_k(\Im\t)^{-1}
(\int_{p}^{z}\!\o-\int_{p}^{z}\!\bo)\Bigr],
\zeile
g_{a_k}\ &=&\ \exp\Bigl[\fr{\p c_1}{2g}\ e_k
(\bt\,(\Im\t)^{-1}\int_{p}^{z}\!\o-\t\,(\Im\t)^{-1}\int_{p}^{z}\!\bo)\Bigr],
\zeile
e_k\ &=&\ (0,\dots,0,1,0,\ldots,0) \qquad\hbox{$k$th unit vector of $\bR^g$}.
\eeql
With these fixed parameters, the transition functions compensate the 
holonomies \gl{mono}, and they are single-valued 
on the {\it boundary\/} of the cut Riemann surface when $c\in\bZ$.

Another convenient choice for the gauge background field strength is 
a $\d$-distributed two-form which has some calculational advantages but 
whose connection cannot be given explicitly.
To construct such a connection one has to use meromorphic one-forms $m$ with 
simple poles at the punctures \cite{farkas}(see Appendix C).
To obtain a $\d$-distributed $U(1)$ field strength one makes use of
\beq
\pa_{\bz}\frac{1}{z}\ =\ \pa_{z}\frac{1}{\bz}\ =\ 2\p\d^2(z).
\eeq
However, since the residues of a globally defined meromorphic one-form always 
sum to zero,
\beq
\sum_{P}\res_{P}(m)\ =\ 0,
\eeq
one has to glue together at least two different one-forms 
by a $U(1)$-transformation.
Let us take a meromorphic one-form $m=\o_{PQ}$ with properties
\beq
\res_{P}(\o_{PQ})\ =\ -\res_{Q}(\o_{PQ})\ =\ 1 \qquad\hbox{and}\qquad
\oint_{a_k}\o_{PQ}\ =\ 0 \quad k=1,\ldots,g 
\eeq
and a holomorphic one-form $\o$.
Let us further choose a disc $\cd$ on the Riemann surface $\cm$ 
such that
\beqx
P \in \cd, \qquad Q \in \cm \backslash \cd.
\eeqx
Now we define two connection one-forms restricted on the two 
sets $\cd$ and $\cm\backslash\cd$ by
\beql\label{deltaA}
A_{c}\ &=&\ \fr{c}{2i}\,(\o_{PQ}-\bo_{PQ}) \qquad\hbox{on}\quad \cd, 
\zeile
A'_{c}\ &=&\ (\o+\bo) \qquad\hbox{on}\quad \cm\backslash\cd.
\eeql
These two connections can be patched together by a $U(1)$-transformation 
$d\L=ig^{-1}dg$ 
which has to satisfy the cocycle condition
\beqx
\oint_{P}d\L\ =\ \oint_{P}(\fr{c}{2i}\,\o_{PQ}-\o+c.c.)\ =\
2\p c\ \in 2\p\bZ. 
\eeqx
The corresponding field strength then reads
\beq\label{deltaF}
F_{c}\ =\ 2\p ic\,\d^2(z{-}z_{P})\,dz\wedge d\bz,
\eeq
and the quantization of the Chern number
follows from the cocycle condition.

\sect{Bosonization in a Gauge Background}
In order to understand the consequences of the gauge moduli 
in the $N{=}2$ string path integral, bosonization of the charged fields 
$\pu{\pm}{}{\m},\b^{\pm},\ga{\mp}$ proves to be very useful.
The bosonized formulation facilitates most explicit calculations.
Let us investigate the changes in the bosonization method due to 
the additional $U(1)$ gauge coupling in the $N{=}2$ string action.
We keep the superconformal gauge but make explicit the
dependence on the gauge field moduli and topology.

It is convenient to first consider only the matter fields
$\pu{\pm}{\da}{}$, $\pu{\pm}{\ua}{}$. 
Since $n=-\fr12$, the index theorem \gl{index} tells us that
a background charge can only come from a non-trivial gauge bundle.
The index $\m$ will be suppressed temporarily.
The part of the action involving a single complex Majorana fermion $\J$ 
reads
\beql\label{matact}
S_\J\ =\ -\fr{1}{8\p}\int\! d^2z\ \Bigl[
	 \pu{-}{\da}{} (\pa_{\bz}+iA_{\bz}) \pu{+}{\da}{}
	&+&\pu{+}{\da}{} (\pa_{\bz}-iA_{\bz}) \pu{-}{\da}{} \zeile
	+\pu{-}{\ua}{} (\pa_z+iA_z) \pu{+}{\ua}{}
	&+&\pu{+}{\ua}{} (\pa_z-iA_z) \pu{-}{\ua}{} \Bigr],
\eeql
and we go to the flat gauge,
\beq
A\ =\ A_z dz+A_{\bz} d\bz\ =\ A_c + h\ 
=\ A_c(z,\bz) + h_z(z)dz + h_{\bz}(\bz)d\bz,
\eeq
with $A_c$ and $h$ given by \gl{deltaA} and \gl{flatA}, respectively.
Of course, we could have gauged $A=0$ completely in any simply-connected
region, and so the (local) bosonization relations should not feel the
presence of the moduli.
Hence, just as for the metric moduli, we shall proceed by first
ignoring $A$ and later accounting for its non-trivial holonomy and
topology.

Because $J_V=*J_A$, we cannot simultaneously bosonize both $U(1)$
gauge currents, as $d\Phi_V=*d\Phi_A$ implies that the two bosons 
would not be mutually local.
We therefore have to make a choice and pick the $U_V(1)$,
which rotates left- and right-movers in the same way.
With $\pu{}{\da}{}$ locally holomorphic and 
$\pu{}{\ua}{}$ locally anti-holomorphic on-shell,
we write
\beq\label{bosonize}
\pu{\pm}{\da}{}(z)\ =\ i\sqrt{2}e^{\pm\f}(z)
\qquad\hbox{and}\qquad
\pu{\pm}{\ua}{}(\bz)\ =\ -i\sqrt{2}e^{\pm\baf}(\bz),
\eeq
with chiral bosons $\f(z)$ and $\bar{\f}(\bz)$ obeying
\beq
\f(z)\ \f(w)\ \sim\ \log(z{-}w) \qquad\hbox{and}\qquad
\bar\f(z)\ \bar\f(w)\ \sim\ \log(\bz{-}\bar w). 
\eeq
The $U(1)$ current $J=J_{z}dz+J_{\bz}d\bz$ becomes~\footnote{
In anti-holomorphic OPEs, anti-radial ordering must
be taken for consistency with conjugation.}
\beql
J_z\ &=&\ \fr12\pu{+}{\da}{}\pu{-}{\da}{}\ =\ -\pa_{z}\f\ =\ -\pa_z\F, 
\zeile
J_{\bz}\ &=&\ \fr12\pu{+}{\ua}{}\pu{-}{\ua}{}\ =\ +\pa_{\bz}\bar\f\ =\ 
		+\pa_{\bz}\F,
\eeql
in accordance with $*J=id\F$, where the full boson 
\beq\label{quantumboson}
\F(z,\bz)\ =\ \f(z) + \bar\f(\bz)
\eeq
satisfies $\pa_{\bz}\pa_z\F=0$ on-shell in regions for vanishing $F$.
It is instructive to check out reality properties.
Under complex conjugation, with the rule $(XY)^*=Y^*X^*$, 
the action \gl{matact} just changes sign~\footnote{
This is because we work with Euclideanized fermions;
in Minkowski space it is invariant.}
if $(\pu{\pm}{\da}{})^*=\pu{\mp}{\ua}{}$,
which implies $J^*=J$ and $\F^*=-\F$ by virtue of $\f^*=-\bar\f$.
We choose all nonchiral bosons to be imaginary.
One has of course also the trivial symmetry of flipping the
$U(1)$ charges and the signs of all bosons (including $A$).

In complete analogy with the standard case, the bosonized action is
\beq\label{bosact}
S_\F\ =\ -\fr{1}{4\p}\int d^2z\Bigl[\pa_{\bz}\F\pa_z\F-2i F_{z\bz}\F\Bigr]\
=\ -\fr{1}{4\p}\int \left[ \fr12 d\F\wedge*d\F + 2F\F\right].
\eeq
The term linear in $\F$ is required in order to agree with
the index theorem,
\beq
4\p[\sharp(\pu{+}{}{})-\sharp(\pu{-}{}{})] = 
i\int dJ = \int d{*}d\F = 2i\int\!d^2z\ \pa_{\bz}\pa_{z}\F =
-2i\int\!d^2z\ F_{z\bz} = 2\int F = 4\p c
\eeq
where $\sharp(.)$ counts the number of zero modes of the 
field in question, 
and we used the Minkowskian equations of motion ($F\to iF$).
Since the gauge freedom allows us to concentrate the
field strength in an arbitrarily small region, we may choose
\beq
-\fr{1}{2\p} \int F\F\qquad\longrightarrow\qquad-c\ \F(P)
\eeq
for some point $P$.
Apparently, the Chern number couples only the zero mode of $\F$.
The latter also controls the conservation of the
charge associated to the global subgroup of $U_V(1)$ in \gl{monos},
\beq
\pu{\pm}{\da\ua}{}\quad\to\quad e^{\pm2\p i\r} \pu{\pm}{\da\ua}{},
\qquad\qquad \f\quad\to\quad\f + 2\p i\r,
\qquad\qquad \bar\f\quad\to\quad\bar\f + 2\p i\r.
\eeq

We take into account the global properties of $\F$.
It is well known \cite{alvarez,verlinde1} 
that the bosonization procedure employs compact bosons, 
which carry integral or half-integral winding numbers 
$m_k,n_k,p_\ell,q_\ell\in\fr12\bZ$ around the homology cycles 
${a_k,b_k,c_\ell,d_\ell}$, respectively,
\beql\label{solimono}
\F(z+a_{k})-\F(z)\ &=\ 2\p im_{k}, \qquad\qquad
\F(z+b_{k})-\F(z)\ &=\ 2\p in_{k}, \zeile
\F(z+c_\ell)-\F(z)\ &=\ 2\p ip_\ell, \qquad\qquad
\F(z+d_\ell)-\F(z)\ &=\ 2\p iq_\ell.
\eeql
Since non-vanishing winding numbers around a puncture are due to terms 
$\F\sim\log(z\bz)$ and $\F\sim\log(z/\bz)$ for a coordinate patch 
centered at the puncture, the bosonic action \gl{bosact} seems to give 
a divergent contribution and thus to suppress $\F$ configurations with
non-trivial puncture windings.
One should, however, regulate these contributions by considering
a genus $g{+}p{-}1$ unpunctured surface close to degeneration,
and take the limit only at the end of the computation.
We separate the multi-valued (solitonic) part of the bosons by writing
\beq\label{fullboson}
\F\ =\ \F_s+\F_q\ =\
(\f_s+\baf_s) + (\f_q+\baf_q).
\eeq
The split is made unique by demanding $\F_s$ to be harmonic.
Again grouping $\tm=(\vec{m},\vec{p})$ and
$\tn=(\vec{n},\vec{q})$, 
the solitonic function is explicitly given by \cite{alvarez,verlinde1}
\beql\label{soliton}
\F_s(P)\ &=&\ 2\p i(\tn-\bT\tm)(T-\bT)^{-1}\int_{P_0}^{P}\O\
	   -\ 2\p i(\tn-T\tm)(T-\bT)^{-1}\int_{P_0}^{P}\bO \zeile
	&=&\ \p \bK (\Im T)^{-1}\!\int_{P_0}^{P}\!\O
	   - \p K (\Im T)^{-1}\!\int_{P_0}^{P}\!\bO
\qquad\hbox{with}\quad K=\tn{-}T\tm \in\fr12\L,
\eeql
on which the action \gl{bosact} evaluates as
\beq\label{solact}
S_{\F}(K)\ =\ \fr{\p}{2} \bK (\Im T)^{-1} K
        -\fr{1}{4\p}\int\!d^2z\ \pa_{\bz}\F_q\pa_{z}\F_q
	-c\,\F(P).
\eeq
One should note that the differential $d\F_s$ is a linear combination 
of holomorphic and antiholomorphic one-forms and therefore not exact, 
in contrast to $d\F_q$.

We must still couple the harmonic gauge connection $h$ with its
nontrivial holonomies to our bosonic system.
The interaction term in the action reads
\beq
S_{\J A\J}\ =\ -\fr{i}{2\p}\int *J\wedge A\
=\ -\fr{i}{2\p}\sum_{i=1}^{g+p-1}\Bigl[
\oint_{\ta_i}*J \oint_{\tb_i}A\ -\ \oint_{\tb_i}*J \oint_{\ta_i}A \Bigr],
\eeq
with the generalized homology cycles
$\ta=(\vec{a},\vec{c})$ and $\tb=(\vec{b},\vec{d})$.
Inserting $A=h$ and $*J=id\F_s$ with winding numbers given in \gl{flatA}
and \gl{solimono}, respectively, we obtain
\beq
S_{\J A\J}(K,C)\ =\ 2\p i \sum_{i=1}^{g+p-1} [\tm_i\tB_i - \tn_i\tA_i]\ 
=\ \p \Bigl[\bK(\Im T)^{-1}C-\bC(\Im T)^{-1}K\Bigr],
\eeq
the intersection number of the one-forms $d\F_s$ and $h$.
This implies that we must extend the bosonized action to
\beq\label{actshift}
S'_{\F}(K,C)\ =\ \fr{\p}{2} [\bK{-}2\bC](\Im T)^{-1}[K{+}2C]
	+2\p\bC(\Im T)^{-1}C
        -\fr{1}{4\p}\int\!d^2z\ \pa_{\bz}\F_q\pa_{z}\F_q
        -c\,\F(P),
\eeq
which is achieved by shifting
\beq\label{join}
d\F_s\ \to\ d\F_s+2*h, \qquad\hbox{that is}\qquad
K\ \to\ K+2C\qquad\hbox{but}\qquad\bK\ \to\ \bK-2\bC.
\eeq
Note that left- and right-movers are shifted in opposite
ways since we have added a real $*h$ to an imaginary $d\F$.\footnote{
Again, the reason is that, in Euclidean space,
$S_\J$ is imaginary where $S_\F$ is real.}
The general form of the bosonized action becomes
\beq
S'_{\F}\ =\ -\fr{1}{4\p}\int \left[ 
\fr12 (d\F+2*h)\wedge(*d\F-2h) + 2*h\wedge h\right] - c\,\F(P)
\eeq
where we dropped $-c\int_{P_0}^P*h$ by choosing $P_0=P$.
Our recipe preserves the form of \gl{soliton} and \gl{solimono}, 
but now with real winding numbers $m_k,n_k,p_\ell,q_\ell\in\bR$. 
The combination of left- and right-moving fermionic correlators
always yields just the bilinears 
\beq
\fr12\pu{\pm}{\da}{}\pu{\pm}{\ua}{}\ =\ e^{\pm\f\pm\baf}\ =\ e^{\pm\F}
\eeq
which are neither real nor $U(1)$ neutral but have well-defined holonomies.
{}From these the bosonized vertex operators are built.

In the bosonized path integral, we are instructed to sum over integral 
solitonic winding numbers for a fixed choice of spin structures, 
as well as to integrate over gauge moduli later on.
The above prescription \gl{join} combines the two.
To make this more explicit, let us perform the standard soliton sum
\beq
f_s(C;D|T)\ =\ \sum_{K\in\fr12\L} 
\exp\Bigl\{-S'_\F[\F_s]+\sum_\ell\a_\ell\,\F_s(P_\ell)\Bigr\}
\eeq
where $D=\sum_\ell\a_\ell P_\ell+cP$ is the divisor of degree $c$ 
given by the vertex and Chern number insertions,
and we use the same symbol for its image under the (generalized)
Abel map, $Z=\int_{P_0}^{P}\O$.
Shifting $K$ according to \gl{join},
the exponent reads
\beql
-S'_\F&&\!\![\F_s]\ +\ \sum_\ell\a_\ell\F_s(P_\ell) \zeile
=&&\ -\fr{\p}{2}[\bK{-}2\bC]{1\over\Im T}[K{+}2C]
-2\p\bC{1\over\Im T}C
+\p[\bK{-}2\bC]{1\over\Im T}D
-\p[K{+}2C]{1\over\Im T}\bar{D} \zeile
=&&\ -\fr{\p}{2}[\bK{-}2\bC{+}2\bar{D}]{1\over\Im T}[K{+}2C{-}2D]
-2\p\bC{1\over\Im T}C-2\p\bar{D}{1\over\Im T}D.
\eeql
Splitting the sum into integral windings plus spin structures and
applying the Poisson resummation formula \cite{alvarez,verlinde1} yields
\beql
f_s(C;D|T)\ &=&\ (\det\Im T)^{1/2}\,
e^{-2\p\Re D{1\over\Im T}\Re D -2\p\bC{1\over\Im T}C}\,
\sum_S\,\Bigl| \Q[S](D-C|T) \Bigr|^2 \zeile
&=&\ (\det\Im T)^{1/2}\,e^{\p F(C,D)}\,
\sum_S\,\Bigl| \Q[S-C](D|T) \Bigr|^2 \quad,
\eeql
where $F(C,D)$ is some real function,
and the sum runs over the (half-integral) spin structures, 
$S\in(\bZ/2\bZ)^{g+p-1}+T(\bZ/2\bZ)^{g+p-1}$.
It is very satisfying that all the gauge moduli hide in the characteristics
of the theta function.
Since the characteristics of the theta function in a fermionic correlator
represent the spin structure of the fermion system involved,
the gauge moduli integral turns into an integral over continuous spin
structures, respectively fermion holonomies, as it should.
Indeed, the chiral fermionic correlator for an arbitrary spin structure
$C$ is proportional to $\Q[C](D|T)$.
Since $S$ corresponds to half-points in the Jacobian torus,
the integral over $C$ is going to cover that torus $2^{2(g+p-1)}$ times.
For a single fermion system (as we have studied here), the final 
gauge moduli integral is simply
\beq
\int\!d^{2(g+p-1)}C\; f_s(C;D|T)\ =\ 2^{3(g+p-1)}\,\det(\Im T)\, 
e^{-2\p\bar{D}{1\over\Im T}D-2\p\bC{1\over\Im T}C},
\eeq
as may be invoked directly by combining the sum and integral
to a regular Gaussian integral.
Due to spectral-flow invariance,
the contribution from the punctures is actually trivial,
as we shall make explicit in the following section.
Their moduli may therefore be dropped altogether,
restricting the puncture winding numbers $p_\ell,q_\ell\in\bZ$.
Thus, multi-loop computations simplify considerably as compared to
the $N{=}1$ string. In particular, there is no need for a GSO projection,
since the sum over spin structures is implicit in the gauge moduli
integration. 

Finally, we will complete the bosonization of the $N{=}2$ string 
by including the Lorentz index $\m$ and the ghost contributions.
The prescription for bosonization of the left-moving charged 
$N{=}2$ fields is given by (now dropping the tildes)~\cite{BKL}
\beql\label{bosgel}
&&\pu{\pm +}{}{}\ =\ \pu{\pm}{}{0} + \pu{\pm}{}{1}, \qquad
\pu{\pm -}{}{}\ =\ \pu{\pm}{}{0} - \pu{\pm}{}{1},
\zeile
&&\pu{\pm +}{}{}(z)\ \pu{\mp -}{}{}(w)\ \sim\ \frac{-4}{z-w},
\zeile
&&\pu{\pm +}{}{}\ =\ 2ie^{+\f^{\pm}}, \qquad
\pu{\mp -}{}{}\ =\ 2ie^{-\f^{\pm}},
\zeile
&&\f^{\pm}(z)\ \f^{\pm}(w)\ \sim\ +\log(z-w),
\zeile
&&\ga{\pm}\ =\ \et{\pm}e^{+\fu{\pm}}, \qquad 
\b^{\mp}\ =\ e^{-\fu{\pm}}\pa\xx{\mp},
\zeile
&&\d(\ga{\pm})\ =\ e^{-\fu{\pm}}, \qquad 
\d(\b^{\mp})\ =\ e^{+\fu{\pm}},
\zeile
&&\fu{\pm}(z)\ \fu{\pm}(w)\ \sim\ -\log(z-w),
\qquad \xx{\mp}(z)\ \et{\pm}(w)\ \sim\ \frac{+1}{z-w},
\zeile
&&\et{\pm}\ =\ e^{-\c^{\pm}}, \qquad \xx{\mp}\ =\ e^{\c^{\pm}}, 
\zeile
&&\c^{\pm}(z)\ \c^{\pm}(w)\ \sim\ +\log(z-w),
\eeql
and correspondingly for the right-movers.
Applying our results obtained for a single complex fermion, 
we bosonize the relevant part of the gauge-fixed action and get
\beql
S_{U(1)}\ =\
-\fr{1}{4\p}\int d^2z&\Bigl\{&\fr{1}{2}(
   \pu{-}{\da}{-}\dvec{\pa}_{\bz}\pu{+}{\da}{+}
  +\pu{-}{\da}{+}\dvec{\pa}_{\bz}\pu{+}{\da}{-})
  -4\b^{-}\pa_{\bz}\ga{+}-4\b^{+}\pa_{\bz}\ga{-} \zeile
&&+4iA_{\bz}[-\fr{1}{4}(
   \pu{-}{\da}{-}\pu{+}{\da}{+}
  +\pu{-}{\da}{+}\pu{+}{\da}{-})
  +\pa_{\bz}(c\tb)+ (\ga{+}\b^{-}-\ga{-}\b^{+})] \zeile
&&+ c.c.\Bigr\} \zeile
\sim\ -\fr{1}{4\p}\int d^2z&\Bigl\{&
   \pa_{z}\F^{+}\pa_{\bz}\F^{+}+\pa_{z}\F^{-}\pa_{\bz}\F^{-}
  -\pa_{z}\Bvf^+\pa_{\bz}\Bvf^+-\pa_{z}\Bvf^-\pa_{\bz}\Bvf^- \zeile
&&+\pa_{z}\Bc^+\pa_{\bz}\Bc^++\pa_{z}\Bc^-\pa_{\bz}\Bc^- \zeile
&&-2iF_{z\bz}[+\F^{+}-\F^{-}-\Bvf^{+}+\Bvf^{-}+c\tb-\bar{c}\bar{\tb}]\zeile
&&+\sqrt{g}\car[\Bvf^++\Bvf^--\fr{1}{2}(\Bc^++\Bc^-)]\Bigr\},
\eeql
where again we retained the gauge field but also the coupling to the
worldsheet curvature.
Decomposing $\F$ and $\Bvf$ into a single-valued ``quantum'' part and
a harmonic solitonic part and inserting the explicit form of the latter, 
we arrive at
\beql\label{uact}
S_{U(1)}\
&=&\ \fr{\p}{2}[\bK^+(\Im T)^{-1}K^+ + \bK^-(\Im T)^{-1}K^-
 - \bar{\ck}^+(\Im T)^{-1}\ck^+ - \bar{\ck}^-(\Im T)^{-1}\ck^-] 
\zeile
&&\quad -c\,
\bigl[\F^{+}{-}\F^{-}{-}\Bvf^{+}{+}\Bvf^{-}{+}c\tb{-}\bar{c}\bar{\tb}\bigr](P)
-4(g{-}1)\bigl[\Bvf^{+}{+}\Bvf^{-}{-}\fr{1}{2}(\Bc^{+}{+}\Bc^{-})\bigr](P')
\zeile
&&\quad -\fr{1}{4\p}\int\!d^{2}z\
   (\pa_{z}\F^{+}_q\pa_{\bz}\F^{+}_q+\pa_{z}\F^{-}_q\pa_{\bz}\F^{-}_q
   -\pa_{z}\Bvf^+_q\pa_{\bz}\Bvf^+_q-\pa_{z}\Bvf^-_q\pa_{\bz}\Bvf^-_q)
\zeile
&&\quad -\fr{1}{4\p}\int\!d^{2}z\
   (\pa_{z}\Bc^{+}\pa_{\bz}\Bc^{+}+\pa_{z}\Bc^{-}\pa_{\bz}\Bc^{-}),
\eeql
where $K^{\pm}$ and $\ck^{\pm}$ contain the winding numbers of the 
solitons $\F_{s}^{\pm}$ and $\Bvf_{s}^{\pm}$, respectively.
The $U(1)$ field strength has been $\d$-distributed at $P$,
and the worldsheet curvature has been concentrated at $P'$.

Due to the ``wrong'' kinetic energy sign of the ghost bosons
in \gl{uact}, their soliton sums appear to diverge.
This phenomenon is well known to also occur in the $N{=}1$ string
and originates in the overcounting of the naive bosonized $(\b,\g)$
path integral due to the picture degeneracy.
Indeed, since two bosons ($\vf$ and $\h$) were used to bosonize
each ($\b,\g$) system, it appears that the solitonic winding lattice
has doubled in dimension.
The remedy \cite{watamura} consists of inserting $g$ projection operators,
which contain a factor of $|\oint{dz\over2\p i}\h(z)|^2$ and fix the 
windings of $\Bvf-\Bc$ to some chosen picture numbers.
The puncture moduli are not affected since we will choose our vertex 
operators in definite pictures. The final result of the amplitude
does not depend on those picture numbers.

Now that we have to deal with four different compact bosons,
the gauge moduli integral is no longer as simple, because
the four solitons share the fractional part $C$ but not the
integral part $K$ of their windings. 
To make sure that all fermions have the same spin structure, however,
one has to modify the bosonized description slightly,
as is also done in the $(1,1)$ string (prior to GSO projection).
Instead of summing the theta-squared of each fermion over the
half-integral spin-structures deriving from its soliton sum,
we first multiply the theta-squares and then sum over common
half-integral spin-structures. The latter sum again combines naturally
with the gauge moduli integral.

The Chern number term in \gl{uact} contains exactly the bosonized form 
of the ghost-extended gauge current $\hat J$, integrated up to $P$.
Abbreviating 
\beq
\hat\f\ =\ \f^{+}-\f^{-}-\vf^{+}+\vf^{-}+c\tb,
\eeq
the weight factor $e^{-S_{g,c}^{fix}}$ in \gl{zgc} is related to the one
at $c=0$ via 
\beq\label{ico}
e^{-S_{g,c}^{fix}}\ =\ e^{-S_{g,0}^{fix}}\;e^{c\hat\F(P)}\
=\ e^{-S_{g,0}^{fix}}\;\SFO(c,P)\;\overline{\SFO}(c,P),
\eeq
with the spectral-flow operator $\SFO$ now defined as~\cite{KL,lechtalk}
\beq\label{sfo}
\SFO(\Q,z)\ =\ \exp\Bigl\{-\Q\int^{z}\!dw\,\hat{J}_w(w)\Bigr\}\
	  =\ \exp\Bigl\{\Q\int^{z}\!dw\,\pa_w\hat{\f}(w)\Bigr\}\
          =\ \exp(\Q\hat{\f}(z)),
\eeq
and $\overline{\SFO}(\Q)=\SFO(-\Q)^*$.
With $z(P)=0$, it should be equivalent to the operator
$\SFO(\Q)$ defined in \gl{sfodef}.
The solitonic piece, $e^{\Q\hat\F_s(P)}$, 
serves to shift the holonomies around $P$. 
The path-integral insertion of $e^{c\hat\F(P)}$ 
originating from \gl{ico} creates a ($c$-fold) instanton 
where the $U(1)$ field strength becomes singular 
and turns a $c=0$ amplitude into one at nonzero value of $c$.
In the fermionic language, it arises from the coupling of the
gauge current $\hat J$ to the background gauge connection $A_c$, 
e.g. \gl{deltaA} and \gl{deltaF}.
Its presence changes not only the fermion zero mode count but also
shifts the total sum of bosonic holonomies from zero to $c$.
Since in the first-quantized formulation there is no a priori
way to compare the normalizations of amplitudes with different values 
of $c$ (or $g$), the ambiguity hiding in the lower bound of the above
integrals (for $c\neq0$) should be absorbed into a new coupling factor 
of $\l^{-c}$, consistent with amplitude factorization.

\sect{Vertex Operators and BRST Cohomology}
The last missing ingredient to calculate string amplitudes
is a set of vertex operators. 
A rather thorough (but not complete) investigation of their 
construction can be found in Ref.\cite{BKL}.
To obtain string vertex operators in explicit form,
bosonization is very helpful.
The bosonization prescription \gl{bosgel} for the susy ghosts 
introduced, in addition to $\vf^\pm$,  a system of 
auxiliary fermions $\et{\pm}$ and $\xx{\pm}$,
which can also be bosonized by $\c^{\pm}$. 
The bosonic $(\fu{\pm},\c^{\pm})$ Fock space is much larger than the
$(\b^{\mp},\g^{\pm})$ Fock space, since the latter is fully represented
once for each value of the picture numbers $\p^\pm$, as measured by
\beq
\P^{\pm}\ =\ \oint[-\pa_{z}\fu{\pm}+\pa_{z}\c^{\pm}].
\eeq

The known procedure to get from one picture to another makes use 
of the two picture-changing operators~\cite{FMS,BKL} 
\beq
\PCO^\pm (z)\ =\ \{ Q , \xi^\pm (z) \}.
\eeq
Inserting the bosonized BRST-current
\beql
J_{BRST}\ &=&\ cT+\et{+}e^{+\fu{+}}\Ge{-}+\et{-}e^{+\fu{-}}\Ge{+}
           +\tc(\pa\F^{-}-\pa\F^{+}) \zeile
         &&+c[\pa cb+\pa\tc\tb 
              -\fr{1}{2}(\pa \fu{+})^{2}-\fr{1}{4}\pa^{2}\fu{+}
              -\fr{1}{2}(\pa \fu{-})^{2}-\fr{1}{4}\pa^{2}\fu{-}
              -\et{+}\pa\xx{-}-\et{-}\pa\xx{+}] \zeile
         &&-4\et{+}e^{\fu{+}}\et{-}e^{\fu{-}}\tb
           +2\pa(\et{-}e^{\fu{-}})\et{+}e^{\fu{+}}\tb
           -2\pa(\et{+}e^{\fu{+}})\et{-}e^{\fu{-}}\tb \zeile
         &&+\fr{3}{4}\pa c(\pa\fu{+}+\pa\fu{+})
           +\tc(\pa\fu{+}-\pa\fu{-}),
\eeql
we find
\beq\label{pco}
\PCO^{\pm}\ =\ c\pa\xx{\pm}+e^{+\fu{\mp}}
(\Ge{\pm}-4\ga{\pm}b \mp 4\pa\ga{\pm}\tb \mp2\ga{\pm}\pa\tb).
\eeq
Since $\PCO^\pm$ are obviously BRST-closed,
they map physical states to physical states.
Furthermore,
\beq
\PCO^+(z)\ \PCO^+(w)\ \sim\ \hbox{regular}, \qquad
\PCO^+(z)\ \PCO^-(w)\ \sim\ \{ Q, \hbox{singular}\} + \hbox{regular},
\eeq
so that picture-changing operators commute on physical states. 
Because $\pa_z\PCO^\pm$ is BRST-exact,
picture-changing operators can be moved at will on the worldsheet,
without changing the correlators of BRST-invariant operators.
In contrast to the $N{=}1$ string, however, $\PCO^\pm$ in \gl{pco}
are {\it not locally invertible\/} and therefore do not immediately 
lead to the usual picture degeneracy of the BRST cohomology.

The spectral-flow operator encountered in the previous section
provides us with another tool to link different pictures.
It is BRST-closed since
\beq
[Q,\SFO(\Q,z)]\ =\ -\Q \Bigl[Q,\int^z \hat J\Bigr]\,\SFO(\Q,z)\
=\ -\Q \Bigl[Q,\int^{z}\{Q,\tb\}\Bigr]\,\SFO(\Q,z)\ =\ 0
\eeq
but not BRST-exact, due to the one in the expansion of the exponential.
In contrast, its derivative is BRST-exact,
\beq
\pa_{z}\SFO(\Q,z) = -\Q J(z)\SFO(\Q,z) = -\Q\{Q,\tb(z)\}\,\SFO(\Q,z)
= -\Q\{Q,\tb(z)\SFO(\Q,z)\}.
\eeq
Thus, $\SFO$ may also be moved around in physical correlators.
The spectral-flow invariance of physical correlators renders 
the gauge puncture moduli integration rather trivial.
For each value of $(\vec{G},\vec{D})$, a puncture twist
may be collected in $\SFO(\Q)\overline{\SFO}(\bar{\Q})$ and
moved away, finally to cancel.
Thus, it suffices to take all vertex operators in the NS sector
and multiply the correlator with the (unit) volume of the
gauge puncture moduli space.
In contrast to $\PCO^\pm$, spectral-flow operators are invertible
by $\SFO(\Q,z)^{-1}=\SFO(-\Q,z)$,
and form an abelian algebra via 
$\SFO(\Q_1)\SFO(\Q_2)=\SFO(\Q_1+\Q_2)$.
Hence, we may insert
\beq\label{SFOinv}
1\ =\ \SFO(\S_i\Q_i,z)\ =\ \prod_i \SFO(\Q_i,z)\ =\ 
\prod_i \SFO(\Q_i,z_i) + \{Q,*\}
\qquad\hbox{with}\quad \sum_i\Q_i=0
\eeq
into any physical correlator.
Moreover, since $\SFO$ clearly commutes with $\PCO^\pm$ and
changes the picture numbers by 
\beq
(\p^+,\p^-)\quad\longrightarrow\quad(\p^++\Q\,,\,\p^--\Q),
\eeq 
it realizes the one-to-one map of spectral flow between physical states 
in all pictures with level $\p:=\p^++\p^-=$ constant.
In particular, NS states at $\p^\pm\in\bZ$ get related to R states 
at $\p^\pm\in\bZ{+}\fr12$.
Nevertheless, the BRST cohomology does change with $\p$, since
there are two {\it inequivalent\/} ways to move by $\d\p^\pm=\fr12$, 
\beq\label{triangle}
\PCO^+ \SFO(-\fr12) - \PCO^- \SFO(+\fr12)\ \neq\ \{ Q,\hbox{something}\}. 
\eeq

To complete the picture, we briefly present the results of our earlier
investigations of the BRST cohomology~\cite{BKL,BL}.
Let us first impose the auxiliary conditions
\beq\label{relative}
b_0 = 0 = \bar b_0 \qquad\hbox{and}\qquad \tc_0 = 0 = \bar{\tc}_0
\eeq
to get rid of the trivial ghost zero mode degeneracy.
The relative BRST cohomology then factorizes into identical
left- and right-moving parts. It turns out that the 
left-moving BRST cohomology in any picture $(\p^+,\p^-)$
is non-zero only at the massless level 
and for the left total ghost number $u=\p+2$. \footnote{
This has yet to be proven for all higher pictures.}
There, however, one finds $2j+1:=\p+3$ different physical 
states~\cite{BL}
\beq
\ket{\p^+,\p^-,q}\ =\ V^{(q)}_{(\p^+,\p^-)} \ket{0},
\qquad q-(\p^+{-}\p^-)/2=-j,\ldots,+j,
\eeq
which are created from the $SL(2)$ invariant neutral NS vacuum 
$\ket{0}$ by vertex operators $V$, whose external momentum $k$
we suppressed.
Here, $q$ denotes the charge under the $U(1)$ factor of the global
$U(1,1)$ Lorentz group.
Since $\PCO^\pm$ are Lorentz singlets but $\SFO(\Q)$ carries charge
$q=\Q$, the proliferation of physical states in higher pictures
goes back to \gl{triangle} whose two terms have different $q$ charge.
This leads to inequivalent words (different $q$) of the same ``length'' 
$\p$ built from $\PCO^\pm$ ($\p{=}1$) and $\SFO$ ($\p{=}0$)
who, acting on $\ket{-1,-1,0}$, create inequivalent physical states 
at picture $(\p^+,\p^-)$.

Thus, the number of physical states seems to grow linearly with
the picture number, but this is not so. In fact, 
it can be shown~\cite{BL} that the states in each $(2j+1)$-plet
are proportional to each other in a nonlocal way, i.e. involving
inverse powers of external momenta.
Combining left- and right-movers again,
this multiplicity is easily understood from the space-time point
of view.\footnote{
O.L. thanks A. Morozov for this suggestion.}
In the lowest picture, $\p=\bar\p=-2$ hence $j=\bar j=0$, 
one has a single scalar state with massless momentum $k$, which
corresponds to a massless scalar field $K(\zu{\pm}{\m})$.
The space-time effect of applying $\PCO^\pm$ or $\overline{\PCO}^\pm$
is to create space-time derivatives, $\pa_{\pm\m}K$,
at level $\p+\bar\p=-3$, represented by the
two doublets $(j=\fr12,\bar j=0)$ and $(j=0,\bar j=\fr12)$.
In this way, any higher derivative of $K$ is represented
in some higher picture, but nothing more.
Furthermore, equating $\PCO$ with space-time derivatives
``explains'' why picture-changing is not (locally) invertible
in the $N{=}2$ string.
Interestingly, the instanton-creation operator, $\SFO(+1)$,
also has a space-time interpretation, but as one of the missing
boost generators which complete the $SO(2,2)$ Lorentz algebra~\cite{BL}.

The canonical vertex operator to be used in our calculational scheme 
for computing string amplitudes (see \gl{zgc}) is
\beq\label{Vcan}
V^{(0)}_{(-1,-1)} \bar{V}^{(0)}_{(-1,-1)} (k)\ =\ 
\tc c\,e^{-\fu{-}-\fu{+}}\ \bar{\tc}\bar{c}\,e^{-\bar\fu{-}-\bar\fu{+}}\  
e^{ik\cdot\zu{}{}},
\eeq
with $k\cdot\zu{}{}=\fr{1}{2}(k^{+\m}\zd{-}{\m}+k^{-\m}\zd{+}{\m})$.
BRST invariance then requires $k^2=0$.
As explained above, any vertex operator in a higher picture can be
constructed by applying an appropriate $\PCO/\SFO$ word to \gl{Vcan}.
In fact, such vertex operators appear when moving some of the
picture-changing insertions in \gl{zgc} and the $\SFO(c)$ insertion
(derived at the end of the previous section) onto punctures.
More generally, using the invariance of physical correlators under
picture-changing and spectral flow \gl{SFOinv}, we may assign
any picture and $q$-charge labels to the vertex operators $V_\ell$ 
in a correlator, as long as the selection rules
\beq
\sum_\ell \p^\pm_\ell = 2g-2 \qquad\hbox{and}\qquad 
\sum_\ell \q_\ell = c
\eeq
are obeyed (using the action $S^{fix}_{0,0}$).
Furthermore, those antighost insertions $\VEV{\m_i|b}$ and
$\langle\o_k|\tb\rangle$ associated with punctures may be contracted with
their vertex operators. A $b$ insertion will replace the $c$ in
\gl{Vcan} by an integral, and a $\tb$ insertion removes the
$\tc$ in front of \gl{Vcan}. 
A list of vertex operators for $\p^\pm=-1,-\fr12,0$ and $q=0,\pm\fr12$
can be found in Ref.\cite{BKL}.

\sect{Conclusions}
We have presented a detailed account of the road leading from the
formal path integral to the actual computation of arbitrary scattering
amplitudes, for the critical $N{=}(2,2)$ string on a flat background.
Using the NSR formulation and including external leg punctures from the 
beginning, this entailed a careful gauge-fixing and discussion of the 
extended $(2,2)$ supermoduli space, with special emphasis
on the gauge moduli arising from the $U(1)$ gauge field of $N{=}2$
worldsheet supergravity. The explicit realization of spectral flow 
and its relation to the gauge moduli was clarified in particular
by employing bosonization in the Abelian gauge background.
Here, solitonic bosons were allowed to carry any real winding number, and 
the resulting spin structure sum found its anticipated extension by the 
gauge moduli integral. Nontrivial gauge bundle topologies were
constructed explicitly and could be reached through the creation of 
worldsheet Maxwell instantons by a local operator generalizing spectral flow.
Finally, we summarized the BRST cohomology of physical states and 
vertex operators, explaining its novel subtleties due to picture changing 
and spectral flow. A new interpretation of picture-changing operators as 
space-time momenta explained their non-invertibility in the $N{=}2$ string.

Having collected all ingredients to evaluate scattering amplitudes at
any loop and instanton number, we stopped just short of that. For once,
the tree-level 3- and 4-point functions have already been
computed in this framework. Furthermore, the results depend on the
choice of allowing for either a single (common left-right) Chern number $c$
or else, two distinct (left and right) Chern numbers $c_L$ and $c_R$.
Since the latter case has been discussed in \cite{berkovits2,vafa,BL}, 
let us close this work with some comments on the former case.

The first nontrivial example is that of the tree-level 3-point function,
where $|c|\le1$. On the sphere, all gauge moduli are trivial and metric
moduli are absent. The two picture-changing operator insertions are moved
to two punctures and change their picture assignments, depending on $c$. 
For $c{=}1$, the instanton-creation operator $\SFO(+1)$ is moved to one
of the vertex operators, increasing its $q$ charge and converting the
picture-assignments to those in the $c{=}0$ sector. Likewise for $c{=}-1$.
The relevant chiral correlators may thus be computed as~\cite{lechtalk,BL}
\beql\label{chiral3}
\tilde A^3_{0,0} \ &=&\ \VEV{ V^{(0)}_{(-1,-1)}(k_1)\,
\oint\!\tb\,V^{(0)}_{(-1,-1)}(k_2)\,\oint\!\tb\,V^{(0)}_{(-1,-1)}(k_3)\;
\PCO^-\PCO^+} \zeile
&=&\ \VEV{ V^{(0)}_{(-1,-1)}(k_1)\,
\oint\!\tb\,V^{(0)}_{(-1,0)}(k_2)\,\oint\!\tb\,V^{(0)}_{(0,-1)}(k_3)} \ =\
-\fr12( k^+_2\cdot k^-_3 - k^-_2\cdot k^+_3) \zeile
\tilde A^3_{0,+1}\ &=&\ \VEV{ V^{(0)}_{(-1,-1)}(k_1)\, 
\oint\!\tb\,V^{(0)}_{(-1,-1)}(k_2)\,\oint\!\tb\,V^{(0)}_{(-1,-1)}(k_3)\;
\PCO^-\PCO^-\SFO(+1)} \zeile
&=&\ \VEV{ V^{(0)}_{(-1,-1)}(k_1)\,
\oint\!\tb\,V^{(0)}_{(-1,0)}(k_2)\,\oint\!\tb\,V^{(+1)}_{(0,-1)}(k_3)}\ =\
- k^+_2\wedge k^+_3 \zeile
\tilde A^3_{0,-1}\ &=&\ \VEV{ V^{(0)}_{(-1,-1)}(k_1)\, 
\oint\!\tb\,V^{(0)}_{(-1,-1)}(k_2)\,\oint\!\tb\,V^{(0)}_{(-1,-1)}(k_3)\;
\PCO^+\PCO^+\SFO(-1)} \zeile
&=&\ \VEV{ V^{(0)}_{(-1,-1)}(k_1)\,
\oint\!\tb\,V^{(-1)}_{(-1,0)}(k_2)\,\oint\!\tb\,V^{(0)}_{(0,-1)}(k_3)}\ =\
+ k^-_2\wedge k^-_3 ,
\eeql
where $\VEV{\ldots}$ denotes the averaging with the action $S^{fix}_{0,0}$.
According to \gl{ico}, left- and right-movers are to be combined using
the same values for $c$, so that the total closed-string tree-level
amplitude (including Maxwell instantons) reads
\beq\label{full3}
A^3_0\ =\ [\tilde A^3_{0,0}]^2 + \l^{-2}[\tilde A^3_{0,+1}]^2
	+ \l^{+2}[\tilde A^3_{0,-1}]^2.
\eeq
This result seems not to be real or neutral under the $U(1)$ factor of the
Lorentz group, but this impression is faulty.
Taking into account the origin \gl{sfo} of the Maxwell
coupling $\l$, together with the fact that $\SFO(\Q)$ carries Lorentz charge
$q=\Q$, one must consider $\l$ as a phase and also assign it a charge $q=1$. 
Obviously, this restores the reality and $U(1,1)$ invariance of $A^3_0$.
One should note, however, that the amplitude \gl{full3} is rather different
from the result of the $c_L\neq c_R$ construction~\cite{BL},
\beq
{A'}^3_0\ =\ [ 2\tilde A^3_{0,0} + \l^{-1} \tilde A^3_{0,+1}
	   + \l^{+1} \tilde A^3_{0,-1} ]^2,
\eeq
which offers the possibility of extending the Lorentz group to $SO(2,2)$.

There remain a number of interesting unresolved issues.
Among them are the structure of one- and higher-loop amplitudes,
a simple proof of the vanishing of $(n{\ge}4)$-point functions (at least
at tree-level) and the extension of this work to open $N{=}2$ string case,
including the interaction of open with closed strings.
The latter is particularly important in order to decide upon the correct 
coupling of self-dual Yang-Mills theory to self-dual gravity in
four dimensions~\cite{ketov}.
We intend to report on these issues in the not too distant future.

\vskip.3in
\subsection*{Acknowledgement}
O.L. acknowledges enlightening discussions with
Gordon Chalmers, Michael Flohr, Amihay Hanany, Sergei Ketov, 
Andrei Losev, Greg Moore, Alexei Morozov, V.P. Nair, Warren Siegel 
and Samson Shatashvili. 
He is also grateful to the Institute for Advanced Study for hospitality
during the final stages of this work.
J.B. would like to thank Peter Adamietz and Klaus J\"unemann for fruitful
discussions.

\begin{appendix}
\newpage
\sect{Conventions}
Spinor Notations:
\beq
\pu{\pm}{}{\m} = \pu{2}{}{\m} \pm i\pu{3}{}{\m}, \qquad
\pu{\pm}{}{\m} = \left(\begin{array}{c}
                         \pu{\pm}{\ua}{\m} \\ \pu{\pm}{\da}{\m}
                         \end{array}\right), \qquad
\bar{\J} = \J\dg \g^0.
\eeq
with $\pu{i}{}{\m}; i =2,3; \m =0,1$ Majorana-Spinors.
$\m$ is the $SO(1,1)$ Lorentz index and $i$ the $SO(2)$ Euclidean 
index of the total $SO(2,2)$ Lorentz group.
Dirac algebra:
\beq
\g^0 = \left(\begin{array}{cc} 0 & -i \\ i & 0 \end{array}\right),\qquad
\g^1 = \left(\begin{array}{cc} 0 & i \\ i & 0 \end{array}\right),\qquad
\g_5 = \left(\begin{array}{cc} 1 & 0 \\ 0 & -1 \end{array}\right).
\eeq
\beql
\{\g^a,\g^b\} &=& 2\h^{ab},\qquad
\g^a \g^b = \h^{ab} + \e^{ab}\g_5, \zeile
\g_b \g^a \g^b &=& 0, \quad\qquad \g^{\m\dg} = \g^0 \g^\m \g^0 \zeile
\h^{ab} &=& \left(\begin{array}{cc} 1 & 0 \\ 0 & -1 \end{array}\right),\qquad
\e^{01} = 1 = \e_{10},
\eeql
Lorentz algebra and its representations:
\beq
\left[L^{ab},L^{cd}\right] = \h^{ad}L^{bc}-\h^{bd}L^{ac}-
                             \h^{bc}L^{ad}+\h^{ac}L^{bd}.
\eeq
\beql
(L^{ab})_r{}^s v_s &=& (\d^a_r \h^{bs} - \d^b_r \h^{as}) v_s,
\zeile
L^{ab} \J &=& \frac{1}{4}\left[\g^a,\g^b\right] \J= \frac{1}{2}\e^{ab}\g_5 \J.
\eeql
Lorentz covariant derivatives:
\beql
D_m v^a &=& \pa_m v^a + \frac{1}{2}\o_{mcd}(L^{cd})_b{}^a v^b
         = \pa_m v^a + \o_m \e^a{}_b v^b,
\zeile
D_m \J &=& \pa_m \J + \frac{1}{2}\o_{mcd}(L^{cd}) \J
        = \pa_m \J + \frac{1}{2}\o_m \g_5 \J,
\eeql
with
\beq
\o_m = \frac{1}{2} \o_{mab}\e^{ab}, \qquad
e_a{}^m \o_m = \e^{nl}(\pa_l e_{na} + i \bgr{-}{l}{}\g_a \gr{+}{n}{}).
\eeq
Light-cone coordinates on the worldsheet:
\beql
x^{\pm} &=& x^0 \pm x^1, \qquad
\pa_{\pm} = \frac{1}{2}(\pa_0 \pm \pa_1), \zeile
\h^{+-} &=& 2, \qquad\h_{-+} = \frac{1}{2},
\eeql
\beq
\g^+ = \left(\begin{array}{cc} 0 & 0 \\ 2i & 0 \end{array}\right),\qquad
\g^- = \left(\begin{array}{cc} 0 & -2i \\ 0 & 0 \end{array}\right),\qquad
\g_5 = \left(\begin{array}{cc} 1 & 0 \\ 0 & -1 \end{array}\right).
\eeq
Wick-rotated ($x^0 \to -ix^0$) coordinates and gamma matrices:
\beqx
x^+ \to z,\quad x^- \to \bz, \quad
\pa_+ \to \pa_z,\quad \pa_- \to \pa_{\bz}, 
\eeqx
\beq
\g^0 \to -i\g^0 = \left(\begin{array}{cc} 0 & -1 \\ 1 & 0 \end{array}\right),
\qquad \g^z = \g^+, \qquad \g^{\bz} = \g^-.
\eeq
Covariant derivatives in complex coordinates:
\beql
D_z\gr{\pm}{\bz}{\ua}&=&\pa_z\gr{\pm}{\bz}{\ua}
                      -\fr{1}{2}\o_z\gr{\pm}{\bz}{\ua},
\qquad
D_{\bz}\gr{\pm}{z}{\da}=\pa_{\bz}\gr{\pm}{z}{\da}
                      +\fr{1}{2}\o_{\bz}\gr{\pm}{z}{\da},
\zeile
D_z\tep{\pm}{\ua}&=&\pa_z\tep{\pm}{\ua}
                      +\fr{1}{2}\o_z\tep{\pm}{\ua},
\qquad\quad
D_{\bz}\tep{\pm}{\da}=\pa_{\bz}\tep{\pm}{\da}
                      -\fr{1}{2}\o_{\bz}\tep{\pm}{\da},
\zeile
D_z\x^{\bz}&=&\pa_z\x^{\bz}+\o_z\x^{\bz},
\qquad\qquad
D_{\bz}\x^{z}=\pa_{\bz}\x^{z}-\o_{\bz}\x^{z}.
\eeql
\beq
\cd_m \ep{\pm}{}=(D_m\mp iA_m)\ep{\pm}{}, \quad
\cd_m \gr{\pm}{n}{}=(D_m\mp iA_m)\gr{\pm}{n}{}
\eeq
One- and two-forms in local complex coordinates:
\beql
\o &=& \o_{z}dz + \o_{\bz}d\bz, \zeile
df &=& \pa_{z} f dz + \pa_{\bz} f d\bz, \zeile
\f &=& \f_{z\bz}dz \wedge d\bz, \zeile
d\o &=& (\pa_{z}\o_{\bz} - \pa_{\bz}\o_{z})dz \wedge d\bz.
\eeql
Hodge star operator: 
\beql
*\o &=& -i\o_{z}dz + i\o_{\bz}d\bz, \zeile
*\f &=& -ig^{z\bz}\f_{z\bz}.
\eeql
Normalized canonical real one-forms:
\beql
\oint_{a_i} \a_j &=& \oint_{b_i} \b_j =\d_{ij}, \zeile
\oint_{b_i} \a_j &=& \oint_{a_i} \b_j =0.
\eeql
Relation with holomorphic one-forms:
\beql
\o_i &=& \a_i + \t_{ij} \b_j, \quad
\bo_i = \a_i + \bt_{ij} \b_j, \zeile
\oint_{a_i} \o_{j} &=& \d_{ij}, \quad
\oint_{b_i} \o_{j} = \t_{ij}, \zeile
\t_{ij} &=& \t_{ji}, \quad
\Im \t_{ij} > 0, \zeile
\b_i &=& (\t-\bt)^{-1}_{ij}(\o_j - \bo_j), \zeile
\a_i &=& (\bt(\bt-\t)^{-1})_{ij}\o_j + (\t(\t-\bt)^{-1})_{ij}\bo_j.
\eeql
Volume forms in complex coordinates $z=x+iy$:
\beql
dz\wedge d\bz &=& -id^2z=-2idx\wedge dy = -2id^2x \zeile
\O &=& \frac{i}{2g}\o_i (\Im\t)^{-1}_{ij}\bo_j
\eeql
Greens and delta functions:
\beql
\int d^2x\d^2(\vec{x}) &=& 1 = \int d^2z\d^2(z), \zeile
\d^2(\vec{x}) &=& 2\d^2(z), \zeile
\pa_{x}^2\pa_{y}^2\ln(r) &=& 2\p\d^2(\vec{x})=2\pa_{z}\pa_{\bz}\ln(z\bz)
                          = 4\p\d^2(z).
\eeql
\beq
\pa_{\bz}1/z = \pa_{z}1/\bz = 2\p\d^2(z).
\eeq
Integration of two closed one-forms:
\beq
\int_{\cm}\Q\wedge\tilde{\Q}=
\sum_{k=1}^{g}[\oint_{a_k}\Q\oint_{b_k}\tilde{\Q}-
               \oint_{b_k}\Q\oint_{a_k}\tilde{\Q}].
\eeq
Lorentz and gauge covariant derivatives (with $n(T^z)=1$):
\beql
\cd^{n,q}_{z}&=&\pa_{z}+n\o_{zz}{}^{z}+2iqA_{z}
              = \pa_{z}+n\o_{z}\ve^{z}{}_{z}+2iqA_{z}\zeile
             &=&\pa_{z}-n\o_{z}+2iqA_{z}, \zeile
\cd^{n,q}_{\bz} &=&\pa_{\bz}-n\o_{\bz}+2iqA_{\bz}.
\eeql
Lorentz covariant derivatives: $D^n=\cd^{n,o}$
\beq
\pa_{m}(\sqrt{g}v^{m})=\sqrt{g}D^{1}_{m}v^m \qquad\Longrightarrow\qquad
\pa_{z}(\sqrt{g}v^{z})=\sqrt{g}(\pa_{z}-\o_z)v^z,
\eeq
for a superconformal gauge with $\gr{\pm}{\bz}{\da},\gr{\pm}{z}{\ua}=0$.
Also, $\overline{\o_z}=-\o_{\bz}$.\\
Adjoint derivatives:
\beql
\int d^2z\sqrt{g}[\bar{\J}^{\l}\cd^{1-\l,q}_{z}\J^{1-\l}]
&=&\int d^2z\sqrt{g}[\bar{\J}^{\l}(\pa_{z}-(1-\l)\o_{z}+2iqA_{z})\J^{1-\l}]
\zeile
&=&-\int d^2z\sqrt{g}
[\overline{(\pa_{\bz}+\l\o_{\bz}+2iqA_{\bz})\J^{\l}}\J^{1-\l}]
\zeile
&=&-\int d^2z\sqrt{g}[\overline{\cd^{-\l,q}_{\bz}\J^{\l}}\J^{1-\l}],
\eeql
\beq
(\cd_{z}^{n,q})\dg=-\cd_{\bz}^{n-1,q}, \qquad
(\cd_{\bz}^{n,q})\dg=-\cd_{z}^{n-1,q}.
\eeq
Index theorem:
\beql
\dim\ker(\cd^{n,q}_{z})-\dim\ker(\cd^{n,q}_{z})\dg &=& (2n-1)(g-1)+2qc,
\zeile
\dim\ker(\cd^{n,q}_{\bz})\dg-\dim\ker(\cd^{n,q}_{\bz}) &=& (2n+1)(g-1)+2qc.
\eeql
\sect{Spectral Flow Operator}
On the modes of the symmetry generators the $\SFO$ has to act by
\beql
\SFO(\Q)L_n \SFO(\Q)^{-1} &=& L_n+\Q J_n+\Q^2\fr{c}{6}\d_n, \zeile
\SFO(\Q)J_n \SFO(\Q)^{-1}&=& J_n+\Q\fr{c}{3}\d_n, \zeile
\SFO(\Q)G^{\pm}_r \SFO(\Q)^{-1}&=& G^{\pm}_{r\mp\Q}.
\eeql
We make an ansatz for a possible $\SFO$ of the following form
\beq\label{ansatz}
\SFO(\Q)=e^{\Q\O}\quad\rightarrow\quad \SFO(\Q)^{-1}=e^{-\Q\O}.
\eeq
Let us start with the spectral flow for the fermionic currents
\beq
\SFO(\Q)G^{\pm}_r \SFO(\Q)^{-1} = G^{\pm}_r+\Q[\O,G^{\pm}_r]
                                       +\frac{\Q^{2}}{2}[\O,[\O,G^{\pm}_r]]+...
\eeq
Considering the power expansion of the fermionic currents for 
an infinitesimal spectral flow with $\Q\ll1$, we obtain 
\beql
G^{\pm}_{r\mp\Q}&=&\oint\frac{dz}{2\pi i}z^{r\mp\Q+1/2}G^{\pm}(z)
                 =\oint\frac{dz}{2\pi i}z^{r+1/2}(1\mp\Q\log(z))G^{\pm}(z)+..
\zeile
&=&G^{\pm}_r\mp\Q\sum_{n}\oint\frac{dz}{2\pi i}z^{-n-1}\log(z)G^{\pm}_{r+n}
 =G^{\pm}_r\mp\Q\sum_{n}c_n G^{\pm}_{r+n}
\zeile
&=&G^{\pm}_r+\Q\sum_{n}c_n[J_n,G^{\pm}_r]=G^{\pm}_r+\Q[\O,G^{\pm}_r]
\eeql
with
\beq
c_n=\oint\frac{dz}{2\pi i}z^{-n-1}\log(z),
\qquad
\O=\oint\frac{dz}{2\pi i}\log(z)J(z).
\eeq
We get a first hint for the exponent $\O$ in \gl{ansatz}.
In order to induce the spectral flow on the $L$- and $J$-modes as well, 
our operator $\O$ has to satisfy the following equations
\beq
[\O,L_{m}]=J_{m},
\qquad
[\O,J_{m}]=\d_{m}.
\eeq
This is only possible if the weird identity 
\beq
c_{n}=\frac{1}{n}\d_{n}
\eeq
is valid. A rather formal calculation gives indeed
\beq
c_n=\oint\frac{dz}{2\pi i}z^{-n-1}\log(z)
   =-\oint\frac{dz}{2\pi i}\frac{1}{n}\partial_{z}(z^{-n})\log(z)
   =\oint\frac{dz}{2\pi i}\frac{1}{n}z^{-n-1}=\frac{1}{n}\d_{n}.
\eeq
Thus, the spectral flow operator given by
\beq
\SFO(\Q)=\exp(\Q\oint\frac{dz}{2\pi i}\log(z)J(z))
\eeq
acts on the modes as well as on the conformal fields in the right fashion.
\sect{Punctured Riemann Surfaces}\label{sieben}
Punctured Riemann surfaces enter the game at the moment one starts to 
consider $p$-string scattering amplitudes. 
The asymptotic incoming and outgoing string states can be conformally mapped 
to points on the compact Riemann surface representing the string 
Feynman diagram. The boundary conditions at these points give the 
quantum numbers of the asymptotic states, for example the momenta. 
These boundary conditions may be encoded by the insertion of vertex operators 
in the path integral over conformal fields living on a punctured 
Riemann surface \cite{dhoker}. 
Instead of treating punctured Riemann surfaces, it is usually more 
convenient to consider the well-known compact surfaces, endowed with 
meromorphic differentials which become singular at specified points. 
These two points of view are equivalent.

Let us first define a standard meromorphic one-form $\o_{PQ}$ having only 
simple poles at the points $P$ and $Q$ with the following properties
\beql
\res_P\o_{PQ} &=& -\res_Q\o_{PQ}=\fr{1}{2\p i}, 
\zeile
\oint_{a_i}\o_{PQ} &=& 0, \qquad i=1,\ldots,g.
\eeql
Such a choice of properties is always possible and determines $\o_{PQ}$ 
uniquely \cite{farkas}. 
In fact all meromorphic differentials have to satisfy 
\beq
\sum_{P\in\cm}\res_{P}m = 0.
\eeq
Every meromorphic one-form with only simple poles with
\beql
\res_{P_\ell}m &=& \q_\ell \qquad\hbox{for}\quad P_\ell\in\cm, 
\quad \ell=1,\ldots,p, \qquad \sum_{\ell}\q_\ell=0,
\zeile
\res_{P}m &=& 0 \qquad\hbox{for}\quad P\in\cm\backslash\{P_1,\ldots,P_p\},
\zeile
\oint_{a_i} m &=& 0 \qquad \hbox{for}\quad i=1,\ldots,g,
\eeql
can be constructed with only the standard one-forms,
\beq
m=\sum_{\ell=1}^{p-1}2\p i(\sum_{k=1}^{\ell}\q_k)\o_{P_\ell P_{\ell+1}}.
\eeq
In this sense the one-forms $\o_{P_\ell P_{\ell+1}}$ for $\ell=1,\ldots,p-1$ 
are extending the standard basis $\{\o_i,\,i=1,\ldots,g\}$ of holomorphic 
one-forms on a genus $g$ Riemann surface 
to a basis for meromorphic one-forms with at most $p$ simple poles. 
This basis is equivalent to a basis of holomorphic one-forms 
on a $p$-punctured surface.

We also have to extend the homology basis of the surface, 
which is dual to the basis of one-forms, 
by adding cycles that surround the points $P_\ell$. 
The appropriate choice for these cycles $c_\ell$, $\ell=1,\ldots,p-1$ is 
given in a way that $c_\ell$ encloses the points 
$P_1$ until $P_\ell$, see Figure 3.
\begin{figure}[t]
\begin{center}
\leavevmode\epsfxsize=8cm
\epsfbox{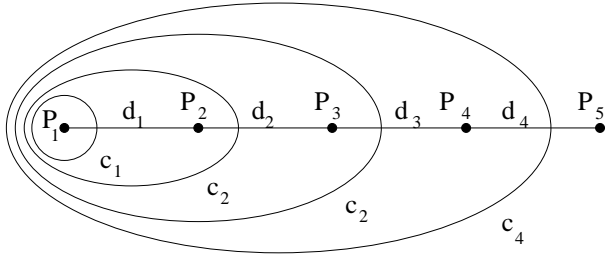}
\caption{Additional cycles for a 5-punctured surface.}
\end{center}
\end{figure}
The line integrals of $\o_{P_\ell P_{\ell+1}}$ along these cycles then 
have the standard form
\beq
\oint_{c_k}\o_{P_\ell P_{\ell+1}}=\d_{\ell k}.
\eeq
Besides the cycles $c_\ell$ we must also introduce the cycles $d_\ell$ 
starting at $P_\ell$ and ending at $P_{\ell+1}$. 

All these structures on a $p$-punctured Riemann surface of genus $g$ are 
identical to the structures of a degenerated compact genus $g+p-1$ surface 
which is obtained 
from the punctured surface by identifying all the points $P_\ell$. 
The cycles $c_\ell$ and $d_\ell$ then correspond to the additional $p-1$ 
$a$- and $b$-cycles one gets for the degenerated compact surface. 

A reasonable definition for the homology basis of a $p$-punctured surface is
\beql
\{a_1,\ldots,a_g,c_1,\ldots,c_{p-1}\}&\equiv&\{a_1,\ldots,a_{g+p-1}\}, \zeile
\{b_1,\ldots,b_g,d_1,\ldots,d_{p-1}\}&\equiv&\{b_1,\ldots,b_{g+p-1}\}.
\eeql
Its dual basis is given by the meromorphic one-forms
\beq\label{extbase}
\{\o_1,\ldots,\o_g,\o_{P_1 P_2},\ldots\o_{P_{p-1}P_p}\}\equiv
\{\o_1,\ldots,\o_{g+p-1}\} 
\eeq
normalized to
\beq
\oint_{a_i}\o_{j} = \d_{ij} \quad\hbox{for}\quad i,j=1,\ldots,g+p-1.
\eeq
The generalized $b$ cycle periods are ($i,j=1,\ldots,g$ here)
\beql
\oint_{b_i}\o_{j} &= \t_{ij}, \qquad\qquad
\oint_{b_i}\o_{P_\ell P_{\ell+1}} &= \r_{i\ell},
\zeile
\oint_{d_\ell}\o_{i} &= \br_{\ell i}, \qquad\qquad
\oint_{d_\ell}\o_{k} &= \s_{\ell k}. 
\eeql
One consequently obtains a $g+p-1$-dimensional period matrix $T$
\beq
\oint_{b_i}\o_j=T_{ij}=\left(\begin{array}{cc} \t&\r\\ \br&\s
                       \end{array}\right)_{ij},
\eeq
where $i,j=1,\ldots,g+p-1$ again.
The degeneration of the Riemann surface by pinching $p-1$ handles to 
one point is reflected in the divergence of the diagonal $\s$ periods
\beq
\s_{\ell\ell} = \int_{P_\ell}^{P_{\ell+1}}\o_{P_\ell P_{\ell+1}}
\eeq
which blow up logarithmically at their endpoints.

We go on to construct bases for the meromorphic differentials 
of the weight 2, $\fr{3}{2}$, and 1, as needed to span the
moduli spaces of the $N{=}2$ string.
Since a one-form $\hat{\o}$ with $p$ simple poles and vanishing 
$a$-periods is uniquely determined for given residues 
it may also be uniquely constructed out of the differentials 
$\o_{P_{\ell}P_{\ell+1}}$.
The Riemann-Roch theorem demands that deg$(\hat{\o})=2g-2$,
which implies that $\hat{\o}$ possesses $2g-2+p$ simple zeroes, 
say at $Q_{0},\ldots,Q_{2g-3+p}$. 
With the help of $\hat{\o}$ we are able to construct the following 
meromorphic differentials \cite{dhoker}:
\begin{itemize}
\item {$3g-3+p$ quadratic differentials:}
\beql
\hat{\o}\o_{i} \qquad &&\hbox{for}\quad i=1,\ldots,g, \zeile
\hat{\o}\o_{Q_{0}Q_{j}} \qquad &&\hbox{for}\quad j=1,\ldots,2g-3+p. 
\eeql
\item {$2g-2+p$ three-half differentials:} \\
For even spin structure there exists no holomorphic but one meromorphic 
half-differential $\k_{Q}$ with a single pole at $Q$, so one has
\beq
\hat{\o}\k_{Q_{j}} \qquad \hbox{for}\quad j=0,\ldots,2g-3+p.
\eeq
For odd spin structure there exists one holomorphic 
half-differential $h$ and in addition a meromorphic 
half-differential $\k_{PQ}$ with single poles at $P,Q$, allowing for 
\beql
&\hat{\o}\k_{Q_{0}Q_{j}}& \qquad\hbox{for}\quad j=1,\ldots,2g-3+p, \zeile
&\hat{\o}h& 
\eeql
\item {$g-1+p$ one-forms:}\\
these were already given in \gl{extbase}, namely
\beq
\o_{i} \qquad \hbox{for}\quad i=0,\ldots,g-1+p.
\eeq
\end{itemize}

Since punctured Riemann surfaces correspond to degenerated compact Riemann 
surfaces of a higher genus, it is possible to generalize the integration rules 
for compact surfaces and holomorphic, respectively antiholomorphic 
differentials $\o,\bo$, given by
\beqx
\int_{\cm}\o\wedge\bo=
\sum_{i=1}^{g}[\oint_{a_i}\o\oint_{b_i}\bo-
               \oint_{b_i}\o\oint_{a_i}\bo].
\eeqx

Let us first consider only two punctures $P,Q$ and the integral 
\beq
\int \o\wedge\bo_{PQ},
\eeq
where $\o_{PQ}$ has the residues $r$ and $-r$ at $P$ and $Q$, respectively. 
For the calculation of the integral above it is again convenient to 
cut the 2-punctured surface along its extended homology basis, see figure 4.
\begin{figure}[t]
\begin{center}
\leavevmode\epsfxsize=6cm
\epsfbox{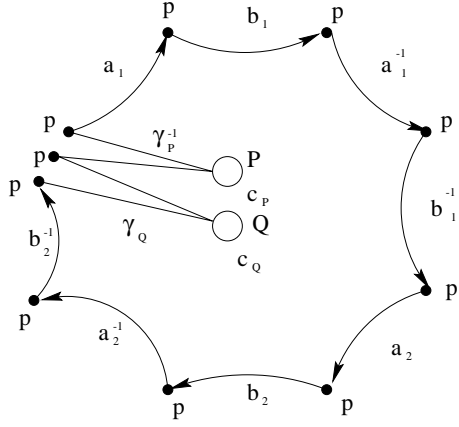}
\caption{Cut 2-punctured Riemann surface.}
\end{center}
\end{figure}
This cut Riemann surface $\cm'$ is simply connected, the differentials 
$\o,\bo_{PQ}$ are (anti)holomorphic and thus closed on $\cm'$. 
By the Poincar\'e lemma $\o,\bo_{PQ}$ are also exact,
\beq
\o=df, \qquad f=\int^{z}_{z_0}\o. 
\eeq
Using the exactness in the integral, one obtains
\beql
\int_{\cm'}\o\wedge\bo_{PQ} =&& \int_{\pa\cm'}f\bo_{PQ} 
=\sum_{i=1}^{g}
\int_{a_i+a_i^{-1}}f\bo_{PQ}+\sum_{j=1}^{g}\int_{b_j+b_j^{-1}}f\bo_{PQ}
\zeile
 &&+\int_{\g_{Q}+\g_{Q}^{-1}}f\bo_{PQ}+\oint_{c_{Q}}f\bo_{PQ}
   +\int_{\g_{P}+\g_{P}^{-1}}f\bo_{PQ}+\oint_{c_{P}}f\bo_{PQ}.
\zeile
=\sum_{i=1}^{g}&&[\oint_{a_i}\o\oint_{b_i}\bo_{PQ}
       -\oint_{b_i}\o\oint_{a_i}\bo_{PQ}] 
       +\oint_{c_{P}}\o\int^{P}_{Q}\bo_{PQ}
       -\int^{P}_{Q}\o\oint_{c_{P}}\bo_{PQ}
\eeql
This is exactly the result which one would expect for the corresponding 
degenerate compact Riemann surface of genus $g+1$. 
The closed path $c_{P}$ corresponds to the additional $a$-cycle and 
the path from $P$ to $Q$ gives the $b$-cycle.

The more general case with two meromorphic differentials may be treated 
in an analogous fashion, 
\beql
\int_{\cm'}\o_{P_{1}P_{2}}\wedge\bo_{P_{3}P_{4}} &=& 
\sum_{i=1}^{g}[\oint_{a_i}\o_{P_{1}P_{2}}\oint_{b_i}\bo_{P_{3}P_{4}}
              -\oint_{b_i}\o_{P_{1}P_{2}}\oint_{a_i}\bo_{P_{3}P_{4}}]
\zeile
&&+\oint_{c_{P_{1}}}\o_{P_{1}P_{2}}\int_{P_{2}}^{P_{1}}\bo_{P_{3}P_{4}}
  -\int_{P_{2}}^{P_{1}}\o_{P_{1}P_{2}}\oint_{c_{P_{1}}}\bo_{P_{3}P_{4}}
\zeile
&&+\oint_{c_{P_{3}}}\o_{P_{1}P_{2}}\int_{P_{4}}^{P_{3}}\bo_{P_{3}P_{4}}
  -\int_{P_{4}}^{P_{3}}\o_{P_{1}P_{2}}\oint_{c_{P_{3}}}\bo_{P_{3}P_{4}}
\zeile
&=&\sum_{i=1}^{g+3}[\oint_{a_i}\o_{P_{1}P_{2}}\oint_{b_i}\bo_{P_{3}P_{4}}
                 -\oint_{b_i}\o_{P_{1}P_{2}}\oint_{a_i}\bo_{P_{3}P_{4}}].
\eeql
This expression becomes logarithmically divergent as soon as two points 
coincide, as is clear from the l.h.s. already.

For a $p$-punctured surfaces and extended cohomology basis 
$\o_{i}$ for $i=1,\ldots,g+p-1$ 
we then get 
\beq
\int_{\cm'}\o_{i}\wedge\bo_{j}=
\sum_{i=1}^{g+p-1}[\oint_{a_i}\o_{i}\oint_{b_i}\bo_{j}
                  -\oint_{b_i}\o_{i}\oint_{a_i}\bo_{j}].
\eeq
\end{appendix}



\end{document}